\documentclass[usegraphicx,useAMS,usenatbib]{mn2e}
\usepackage[T1]{fontenc}
\usepackage{verbatim}
\usepackage[normalem]{ulem} 
\usepackage{amsfonts,amsmath,amssymb} 
\usepackage{lipsum,multicol}
\usepackage{times}
\usepackage[italic]{mathastext} 
\usepackage{enumitem}
\usepackage[colorlinks,linkcolor=blue,citecolor=blue,urlcolor=blue]{hyperref}
\usepackage{xcolor,colortbl}

\usepackage{multirow}

\usepackage{color}
\usepackage{upgreek}
\usepackage{flushend}
\usepackage{tikz}

\usepackage{todonotes}

\title[Aging Halos: Conditional Statistics of Massive Halos]{Aging Halos: Implications of the Magnitude Gap on Conditional Statistics of Stellar and Gas Properties of Massive Halos}

\author[A.~Farahi, et al.]{
Arya~Farahi$^{1,2}$\thanks{corresponding author: aryaf@umich.edu},
Matthew~Ho$^{1}$, and
Hy~Trac$^{1}$.
\\
\\
$^{1}$ McWilliams Center for Cosmology, Department of Physics, Carnegie Mellon University, Pittsburgh, PA 15312, USA\\
$^{2}$ Michigan Institute for Data Science, University of Michigan, Ann Arbor, MI 48109, USA\\
}

\begin{document}
\date{\today}
\pagerange{\pageref{firstpage}--\pageref{lastpage}} \pubyear{2018}
\maketitle
\label{firstpage}

\begin{abstract}
Cold dark matter model predicts that the large-scale structure grows hierarchically. Small dark matter halos form first. Then, they grow gradually via continuous merger and accretion. These halos host the majority of baryonic matter in the Universe in the form of hot gas and cold stellar phase. Determining how baryons are partitioned into these phases requires detailed modeling of galaxy formation and their assembly history. It is speculated that formation time of the same mass halos might be correlated with their baryonic content. To evaluate this hypothesis, we employ halos of mass above $10^{14}\,M_{\odot}$ realized by TNG300 solution of the IllustrisTNG project. Formation time is not directly observable. Hence, we rely on the magnitude gap between the brightest and the fourth brightest halo galaxy member, which is shown that traces formation time of the host halo. We compute the conditional statistics of the stellar and gas content of halos conditioned on their total mass and magnitude gap. We find a strong correlation between magnitude gap and gas mass, BCG stellar mass, and satellite galaxies stellar mass, but not the total stellar mass of halo. Conditioning on the magnitude gap can reduce the scatter about halo property--halo mass relation and has a significant impact on the conditional covariance. Reduction in the scatter can be as significant as 30\%, which implies more accurate halo mass prediction. Incorporating the magnitude gap has a potential to improve cosmological constraints using halo abundance and allows us to gain insight into the baryon evolution within these systems.
\end{abstract}

\begin{keywords}
  galaxies: clusters: general,
  galaxies: clusters: statistics,
  galaxies: haloes,
  galaxies: stellar content
\end{keywords}

\section{Introduction} 
\label{sec:introduction}

The most massive gravitationally bound objects in the Universe, i.e. clusters of galaxies, contain vast amounts of dark matter (DM), plus baryonic matter in two phases, a hot gas component ($ > 10^6$~K) and a cold stellar phase. Population statistics of these systems as a function of their baryonic observables are a valuable probe of cosmological parameters \citep{Allen:2011review,Weinberg:2013,Huterer:2018}.  A key observable is the number density of clusters as a function of baryonic observables. However, there is a gap between what theory predicts and observational quantities. The theory predicts the number density of dark matter halos as a function of their mass and redshift \citep[{\sl e.g.},][]{Jenkins:2001, Evrard:2002, Tinker:2008, Murray:2013}, while halo mass is not directly observable. The main difficulty is how to infer the mass of these systems from the baryonic observables, as the dark sector is not directly accessible to us. To build a precise and accurate model of halo mass -- baryonic observables, we need a comprehensive understanding of gravity and the astrophysics.

The key quantifiable observable is a probabilistic relation between the mass of dark matter halo and its baryonic observables \citep{Mulroy:2019}. Assuming gravitational collapse and violent relaxation \citep{LyndenBell:1967} under the approximation of spherical symmetry, self-similar solutions emerge for both collisionless and collisional fluids \citep{Bertschinger:1985}. But this simple model ignores non-spherical and non-gravitational effects such as hierarchical mergers driven by large-scale filaments and feedback by star formation, supernovae, and active galactic nuclei (AGN). A central focus of recent observations and modern simulations has been on understanding these non-gravitational effects and feedback processes, which leads to a comprehensive understanding of the ways in which the galaxies are forming and evolving as a function of their host halo mass and redshift \citep{McCarthy:2017, Pillepich:2018scaling, Mulroy:2019}. While the first principle approaches have been made progress in making more realistic galaxy population, empirical approaches are complementary in gaining insight into the physics of these systems. One empirical approach that enhances our understanding of these physical processes is to look for a secondary physical parameter, beyond mass and redshift, that is correlated with the baryonic observables.

The relation between the mass of the most massive halos, $M \gtrsim 10^{14}\,[M_{\odot}]$, and its baryonic observables, i.e. the mass--observable relation, can be modeled with a power-law relation and a log-normal distribution \citep{Kaiser:1991}. In the context of halo population statistics, the log-normal model is a mathematically convenient form to work with \citep{Evrard:2014}, and is supported by theoretical arguments, simulations, and observational studies. For example, in the context of modeling star formation, a log-normal shape for final stellar mass naturally emerges when random factors govern the evolution of the system \citep[{\sl e.g.},][]{Larson:1973, Zinnecker:1984}. In the past decade, observations \citep{Pratt:2009, Reichert:2011, Mahdavi:2013, Lieu:2016xxl, Mantz:2016-relaxedIII, Mantz:2016WtG-V, McClintock:2019, Golden-Marx:2019, Mulroy:2019} and simulations \citep[][]{Evrard:2008, Stanek:2010, Truong:2018, Farahi:2018bahamas, Bradshaw:2019} have employed the power-law, log-normal model and extensively studied and quantified the mean and scatter of mass--observables relation. These studies found that the a power-law model with a log-normal scatter is a sufficient model to describe integrated properties of the most massive systems. We will also employ this model to describe the stellar and gas properties of our sample.

As a consequence of the self-similar model, it is commonly assumed that halo properties primarily scale with mass and secondarily evolve with redshift \citep{Kaiser:1991}. While there are theoretical attempts to model the mass--observable scatter \citep[{\sl e.g.},][]{Afshordi:2002, Okoli:2016}, these models implicitly assume that the scatter in a halo property at a given halo mass and redshift is intrinsic and irreducible. Recent simulations and observational studies have challenged this assumption and shown that including the formation time or a proxy of formation time reduces the scatter in the brightest central galaxy (BCG) stellar mass--halo mass relation \citep{Matthee:2017, GoldenMarx:2018, Golden-Marx:2019, Bradshaw:2019}. The hierarchical formation scenario implies that the formation time, besides the halo mass, is another key quantity that controls the final stellar mass of BCG \citep{GoldenMarx:2018}. This postulation is supported by the the observational data \citep{GoldenMarx:2018, Golden-Marx:2019}, and simulations of both the Milky Way size halos \citep{Matthee:2017} and cluster size halos \citep{Bradshaw:2019}. A consequence of this finding is that the scatter about halo property--halo mass relation (HPHM) could be partially explained. We extend this argument and postulate that halo formation time contains information about the final state of the hot gas and the stellar content of dark matter halos. This postulation implies that conditioning halo properties on a proxy of formation time reduces the scatter about HPHM relation. Establishing such a correlation, if exists, deepens our understanding of physical processes that controls the final baryon content of dark matter halos. Our primary aim is to evaluate this postulation using a set of halos derived from hydrodynamics simulations.

The magnitude gap, the magnitude difference between the magnitude of the brightest and $n^{th}$ brightest galaxy, is a strong indicator of the formation time \citep{Jones:2003, DOnghia:2005, Dariush:2010, Deason:2013, Kundert:2017, GoldenMarx:2018, Vitorelli:2018}.  A simple dynamical argument can explain the correlation between formation time and the magnitude gap. A massive satellite galaxy merges with the central galaxy, the galaxy that sits at the center of the host dark matter halo, over a dynamical timescale. The earlier a halo is formed, the more likely that the central galaxy cannibalizes nearby massive satellite galaxies, leaving only less massive and less luminous survivors with a larger magnitude gap. This result implies that the magnitude gap contains information about halo properties that supplements the information provided by halo mass. We show that conditioning on magnitude gap indeed reduces the scatter about the mean HPHM. Another goal of this work is to quantify the impact of conditioning on magnitude gap on HPHM parameters -- the mean relation, the scatter, and the halo property pair covariance. The set of halo properties that we study in this work includes total stellar and gas mass, count of satellite galaxies, and stellar mass of the central and satellite galaxies.

Using a set of halos derived from the TNG300 solution of the IllustrisTNG project \citep{Pillepich:2018}, we study the conditional statistics of halo properties at given halo mass and magnitude gap. We find that conditioning on magnitude gap changes the scaling relation parameters and the entire covariance matrix, scatter and the correlation coefficients. This conditioning allows us to study halos of similar mass and age and identify quantities which are mainly correlated due to the age, like the number of satellite galaxies and BCG stellar mass. Conditioning on magnitude gap makes some correlations stronger because of elimination of additional scatter due to the mixing of halos of different ages, for example the correlation between satellites stellar mass and the total stellar mass. These predictions provide another set of observables which can be tested via observational data.

Reducing the scatter does not only clue in on galaxy formation and astrophysics but also has an impact on the cosmological parameters estimation. This scatter parameter controls how accurately the mass of these systems can be estimated given a set of properties and governs the accuracy of a cosmological analysis \citep{SDSS:2018cosmology, Farahi:2019}.  Additionally, reducing the scatter can lessen the impact of systematics -- such as selection bias and non-Gaussianity in the mass--observable relation -- on a cosmological analysis. Therefore, identifying secondary physical parameters that are correlated with the halo properties at a given halo mass and redshift is of extreme interest. We dub such a parameter a secondary explanatory variable.

Before continuing, we have to distinguish between halo properties and cluster observables. In the literature, these two terms, while drastically different, are frequently used interchangeably. Halo properties are the intrinsic properties of dark matter halos that can be studied with simulations, while cluster observables are quantities measured on the projected sky. The projection effect can induce a non-trivial correlation between different observables and add additional bias and scatter \citep{Costanzi:2018projection}. Therefore, before drawing any definite conclusion from an observational study about the intrinsic properties of dark matter halos, we first need to understand the key observational systematics such as the projection effect and the mapping between mass and light. In this work, we only focus on the halo properties and the mapping from halo properties to cluster observables is a subject of future work.

The structure of this paper is as follows. In Section \ref{sec:data}, we introduce the simulation and the sample employed in this work. In Section \ref{sec:population}, we describe the population model and regression algorithm employed to obtain the halo property--halo mass relation, with results presented in Section \ref{sec:results}. In Section \ref{sec:discussion}, we discuss the implications of the magnitude gap in a scaling relation analysis, observational systematics and challenges, and speculate about other potential secondary explanatory parameters. Finally, we summarize our results and conclude in Section \ref{sec:conclusion}. We include one appendix, Appendix \ref{app:covariance-splitting}, where we propose a model of the property covariance matrix that quantifies the impact of the secondary explanatory variable.

{\sl Notation and Cosmology:} The TNG universe is a flat $\Lambda$CDM cosmology with $\Omega_{\Lambda,0} = 0.6911$, $\Omega_{\Lambda,0} = 0.3089$, $\Omega_{b,0} = = 0.0486$, $\sigma_8 =
0.8159$, $n_s = 0.9667$, and $H_0=67.74$ km s$^{-1}$ Mpc$^{-1}$. Distances and masses, unless otherwise noted, are defined as physical quantities with this choice of cosmology, rather than in comoving coordinates. We denote the mass inside spheres around the cluster center as $M_{200}$, corresponding to an overdensity of $200$ times the critical density of the TNG universe at redshift zero, $\rho_{\rm crit}(z=0) = 3 H_0^2 / 8 \pi G$. We denote logarithm base $e$ as $\ln$. Unless otherwise noted, we report the scatter in $\ln$-base.

\section{The Sample} \label{sec:data}

In this section, we describe the halo sample and halo quantities employed in this work. The sample and measured quantities are derived from the TNG300 solution of the IllustrisTNG project \footnote{\url{http://www.tng-project.org/data/}}. Our sample consists of a subset of data that is provided publicly by the IllustrisTNG team \citep{ Nelson:2018, Pillepich:2018, Springel:2018, Marinacci:2018, Naiman:2018}.

The TNG solutions are based on the deformable-mesh code, Arepo \citep{Springel:2010EMesh, Weinberger:2019} that incorporate AGN feedback, cooling star-formation, and supernovae feedback with cosmological parameters listed in Table~\ref{tab:samples}. Free parameters in the feedback model were tuned to match some observational quantities, such as the evolution of the cosmic star formation rate density \citep{Vogelsberger:2014}. 

\subsection{Halo Properties}

We utilize a halo and subhalo catalog products of the TNG300 simulations that provided by the IllustrisTNG team \citep{Pillepich:2018}. Halos are identified using a ``friends-of-friends'' percolation algorithm. In this work, we employ a mass-limited sample with $M_{200} > 10^{14}\,[M{\odot}]$ at redshift $z = 0$. Our sample consists of 280 halos with maximum halo mass of $1.5 \times 10^{15}\,[M_{\odot}]$ and  nine properties measured within a spherical region. These quantities are listed in Table \ref{tab:sim-results}. These are spherically integrated quantities measured using the minimum of the local gravitational potential as the halo center, and any sub-halos and particles that lie outside the characteristic radii $R_{200}$ are discarded. The measured properties include a halo mass, $M_{200}$, aggregate stellar mass, $M_{\rm star}$, total gas mass, $M_{\rm gas}$, and the sum of gas and stellar mass, $M_{b}$, measured within a spheres of radius $R_{200}$.

The total stellar and gas mass of a system is hard to measure empirically. We thus include effortlessly measurable quantities, such as the stellar mass of the brightest central galaxy (BCG), $M_{\rm star, BCG}$, the stellar mass of the satellite galaxies above a magnitude limit, $M_{\rm star, SAT}$, total number of satellite galaxies above a magnitude limit, $N_{\rm gals}$ and r-band magnitude gap measures. r-band magnitudes are computed by summing up the luminosity of all the stellar particles of the subhalo in SDSS equivalent r-band. Stellar mass of the satellite and BCG is the sum of the mass of the member stellar particles, but restricted to stars within the radius at which the surface brightness profile drops below the limit of $20.7\,{\rm mag}\,{\rm arcsec}^{-2}$ in a $K$-band. To compute the stellar mass of satellite galaxies and $N_{\rm gals}$, we select only satellite galaxies with r-band magnitude $M_r < -19$. M12 (M14) is defined as the difference between the r-band magnitude of the brightest and the second (fourth) brightest member galaxies within the $R_{200}$ radii.

\begin{figure}
     \centering
     \includegraphics[width=0.49\textwidth]{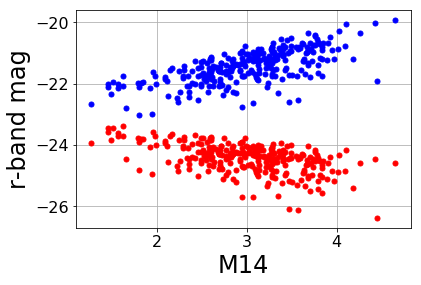}
     \caption{Distribution of r-band absolute magnitude of the BCG and the fourth brightest member galaxy versus M14. The BCGs are shown in red and the fourth brightest galaxy members are shown in blue.}
     \label{fig:M14-dist}
\end{figure}

Figure \ref{fig:M14-dist} shows the distribution of the BCG r-band magnitude and the fourth brightest galaxy r-band magnitude as a function of M14. From this figure, it is evident that there is a strong correlation between the magnitude gap and the brightness of the BCG. We will discuss this in more detail in Sections \ref{sec:results} and \ref{sec:discussion}.

\begin{table*} 
	\begin{center}
    \caption{The cosmological parameters assumed to perform the TNG300 simulations of the IllustrisTNG project. It assumes a flat $\Lambda$CDM cosmology, thus $\Omega_\Lambda = 1 - \Omega_m$. $N_{\rm sam}$ is our sample size, subset of halos with $M_{200} > 10^{14}\, M_{\odot}$. $L$ is the comoving simulation cube length.} \label{tab:samples}
		\begin{tabular}{|l | c c c c c c c c |} 
            \hline 
			Simulation & $N_{\rm sam}$ & $L$ (Mpc) & $\Omega_m$ & $\Omega_b$ & $f_b$ & $\sigma_8$ & $n_s$ & h\\
            \hline
            TNG300-1& 280 & 302.6 & 0.3089 & 0.0486 & 0.157 & 0.8159 & 0.9667 & 0.6774 \\ \hline
            \end{tabular}
	\end{center} 
\end{table*}

\begin{table*} 
\caption{Halo properties employed in this work. These properties are spherically integrated quantities within $R_{200}$ of the host halo. These quantities are derived from the TNG300 solution at redshift $0$. \citet[][and references therein]{Pillepich:2018} provide details of the IllustrisTNG project and how these quantities are measured. The horizontal line above M12 splits halo properties into two groups. Halo properties above the line are the elements of vector $\textbf{S}_p$, i.e. the dependent variables. The quantities below the line are employed as a secondary explanatory properties $X_2$, except $M_{\rm star, BCG}$ which is used both in $S_p$ and $X_2$. See Section \ref{sec:population} for the definition of $S_P$ and $X_2$.}   \label{tab:halo-quantities}
	\begin{center}
		\tabcolsep=0.8mm
		\begin{tabular}{ | l | l | l | }
            \hline
            Element & Unit & Description  \\
            \hline
$M_{\rm halo}$ & $10^{14}\, M_{\odot}$ & Total mass of halo within $R_{200}$ \\
$M_{\rm gas}$ & $10^{13}\, M_{\odot}$ & Total mass gas particles within $R_{200}$ of the host halo. \\
$M_{\rm star}$ & $10^{12}\, M_{\odot}$ & Total mass stellar particles within $R_{200}$ of the host halo. \\
$M_{\rm star, BCG}$ & $10^{11}\, M_{\odot}$ & Sum of the mass of the member stellar particles for the brightest galaxy in r-band$^{1}$. \\ 
$M_{\rm star, SAT}$ & $10^{11}\, M_{\odot}$ & Sum of the mass of the member stellar particles of galaxies with $M_{r} < -19$ except the brightest galaxy in r-band$^1$. \\ 
$M_{\rm b}$ & $10^{13}\, M_{\odot}$ & Sum of the gas mass and the stellar mass within $R_{200}$ of the host halo. \\
$N_{\rm gals}$ & none & Count of galaxies within $R_{200}$ of the host halo with $M_{r} < -19$. \\ \hline
M12  & mag & The r-band magnitude difference between the brightest and the second brightest galaxy within $R_{200}$ of the host Halo. \\
M14  & mag & The r-band magnitude difference between the brightest and the fourth brightest galaxy within $R_{200}$ of the host Halo. \\
            \hline
        \end{tabular}
            \item[$^1$] Restricted to stars within the radius at which the surface brightness profile drops below the limit of 20.7 mag arcsec$^{-2}$ in a $K$-band.
	\end{center}
\end{table*}

\section{Population Statistics} \label{sec:population}

The main focus of this work is on the conditional statistics of integrated Baryonic properties of dark matter halos. Throughout this paper, we use the term ``halo properties'' to describe integrated Baryonic properties of dark matter halos; and the term ``secondary explanatory variable'' is used to specify a property of dark matter halo to explain the scatter about halo mass--halo property relation. We define log-space halo properties $\textbf{s}_p \equiv \ln(\textbf{S}_p)$, where $\textbf{S}_p$ is the vector of halo properties given in Table~\ref{tab:halo-quantities}. The conditional probability distribution of halo properties is specified with $P(\textbf{s}_p \, | \, M, X_2)$ where $M$ is the halo mass and $X_2$ is a secondary explanatory variable.  To model this conditional statistics, we employ a multi-variate log-normal likelihood.  This model and notation has a similar spirit to the model and notation employed in \citet{Evrard:2014}.

The log-normal model, besides its mathematical tractability, is justified theoretically and verified through hydrodynamical simulations \citep[{\sl e.g.},][]{Evrard:2008, Stanek:2010, Truong:2018, Farahi:2018bahamas}. For example, \citet{Farahi:2018bahamas} illustrate that the integrated stellar mass and gas mass of halos extracted from the BAHAMAS and MACSIS hydrodynamical simulations follow a log-normal form. Our results are also in good agreement with the log-normal model. In Section \ref{sec:results}, we demonstrate that the scatter of halo properties about mean relation is consistent with a log-normal form, supporting the choice for our population inference model. 

\subsection{Regression model} \label{sec:regression}

We employ a linear model between log of a halo property and log of halo mass and a secondary explanatory variable. Table \ref{tab:halo-quantities} lists halo properties and explanatory variables employed in this work. Assuming a log-normal model, the likelihood of having logarithmic halo property $s_{p} \equiv \ln(S_{p})$ is
 \begin{equation} \label{eq:Likelihood}
P(s_{p} \, | \, M_{\rm halo}, X_2) \ \propto
 \exp\left\{ - \frac{ \left( s_{p} - \langle s_{p} \, | \, M_{\rm halo}, X_2 \rangle \right)^2}{ 2 \sigma_{\rm int}^2 } \right\} .
\end{equation}
where $\sigma_{\rm int}$ is the intrinsic scatter and $X_2$ is the secondary explanatory property. We test and verify the log-normal assumption in Section \ref{sec:log-normal}. In our model, the log-mean relation follows a linear relation 
\begin{equation} \label{eq:log-mean}
 \langle s_{p} \, | \, \ln M_{\rm halo}, X_2 \rangle \ = \ 
  \pi  + \alpha \, \ln M_{\rm halo} + \beta \, X_2, 
\end{equation}
where $\alpha$ and $\beta$ are the slopes of log mass and secondary explanatory variable respectively and $\pi$ is the logarithmic intercept at redshift zero. While we only show the results of the above linear model, to assess the sensitivity of our results to this linear choice, we allow higher-order terms in our scaling relation model.

We implement the above regression model in \texttt{PyMC} \citep{PyMC:2016} which allows us to sample the posterior distribution of the model parameters. We employ a set of weakly informative priors, a wide Gaussian distribution for the slope and the normalization and a wide truncated Gaussian distribution for the scatter parameter. These priors are specified in Table \ref{tab:priors}. Our results are insensitive to the choice of these priors.

\begin{table}
\caption{Model Priors. } \label {tab:priors}
	\begin{center}
		\tabcolsep=0.8mm
		\begin{tabular}{ | l | l | l | }
            \hline
            Parameter & Prior \\
            \hline
$\pi$ & $\mathcal{N}(\mu = 0, \sigma = 10^6)$\\
$\alpha$ & $\mathcal{N}(\mu = 0, \sigma = 10^6)$ \\
$\beta$ & $\mathcal{N}(\mu = 0, \sigma = 10^6)$ \\
$\sigma_{\rm int}$ & $T\mathcal{N}(\mu = 0, \sigma = 10^4; a=0)$\\ 
            \hline
        \end{tabular}
	\end{center}
\end{table}

While slope and scatter of the mean relations may be scale-dependent and/or evolve with redshift \citep[{\sl e.g.},][]{LeBrun:2017, Truong:2018, Farahi:2018bahamas}, at the mass scales that are considered in this work the scale-dependence is negligible. The current observational data are yet not rich enough to model these effects. We test this in two ways. First, we allow higher-order terms in our scaling model, including $\ln(M)^2$ and $X_2^2$. The coefficients of these terms are consistent with zero, and including them does not have a statistically significant impact on our results and conclusions. Second, we test the mass dependence by splitting both samples into two non-overlapping subsets at the pivot mass, and find no evidence of scale dependence in the posterior scaling relation parameters, which is expected for the high end of the mass function \citep{Farahi:2018bahamas}.

\subsection{Halo Property Pair Covariance} \label{sec:halo-prop-corr-model}

 The term ``halo property pair covariance'' is employed to describe the covariance between a pair of halo properties at fixed halo mass. We employ the estimator
\begin{equation} \label{eq:r-estimator}
{\rm COV(s_a,s_b)} = \frac{1}{n-1}\sum\limits_{i=1}^{n} ~ \delta s_{a,i} ~ \delta s_{b,i}. 
\end{equation}
to compute halo property pair covariance. $\delta s_{a,i}$ is the residual of log property $s_a$ for halo $i$ about the mean HPHM relation, 
\begin{equation}\label{eq:normedResidual}
    \delta s_{a,i} =  s_{a,i} - \langle s_{a,i} \, | \, M_{\rm halo}, X_2 \rangle \,. 
\end{equation}
Consequently, the corresponding property pair correlation coefficient is 
\begin{equation} \label{eq:rab}
r_{a, b} = \frac{{\rm COV(\delta s_a, \delta s_b)} }{\sigma_a \, \sigma_b}, 
\end{equation}
where $\sigma_a = \sqrt{ {\rm COV(\delta s_a,\delta s_a)}}$ is the intrinsic scatter in property $a$ at fixed halo mass and secondary parameter, and similarly for property $b$. To compute the statistical uncertainty, we employ a bootstrap approach. We bootstrap the residuals 1,000 times and compute the correlation coefficient for each bootstrapped sample. We report the mean and standard deviation of the bootstrapped samples. The results of halo property pair covariance are presented in Section \ref{sec:halo-prop-corr}.

\section{Results} \label{sec:results}

\begin{figure*}
     \centering
     \includegraphics[width=0.98\textwidth]{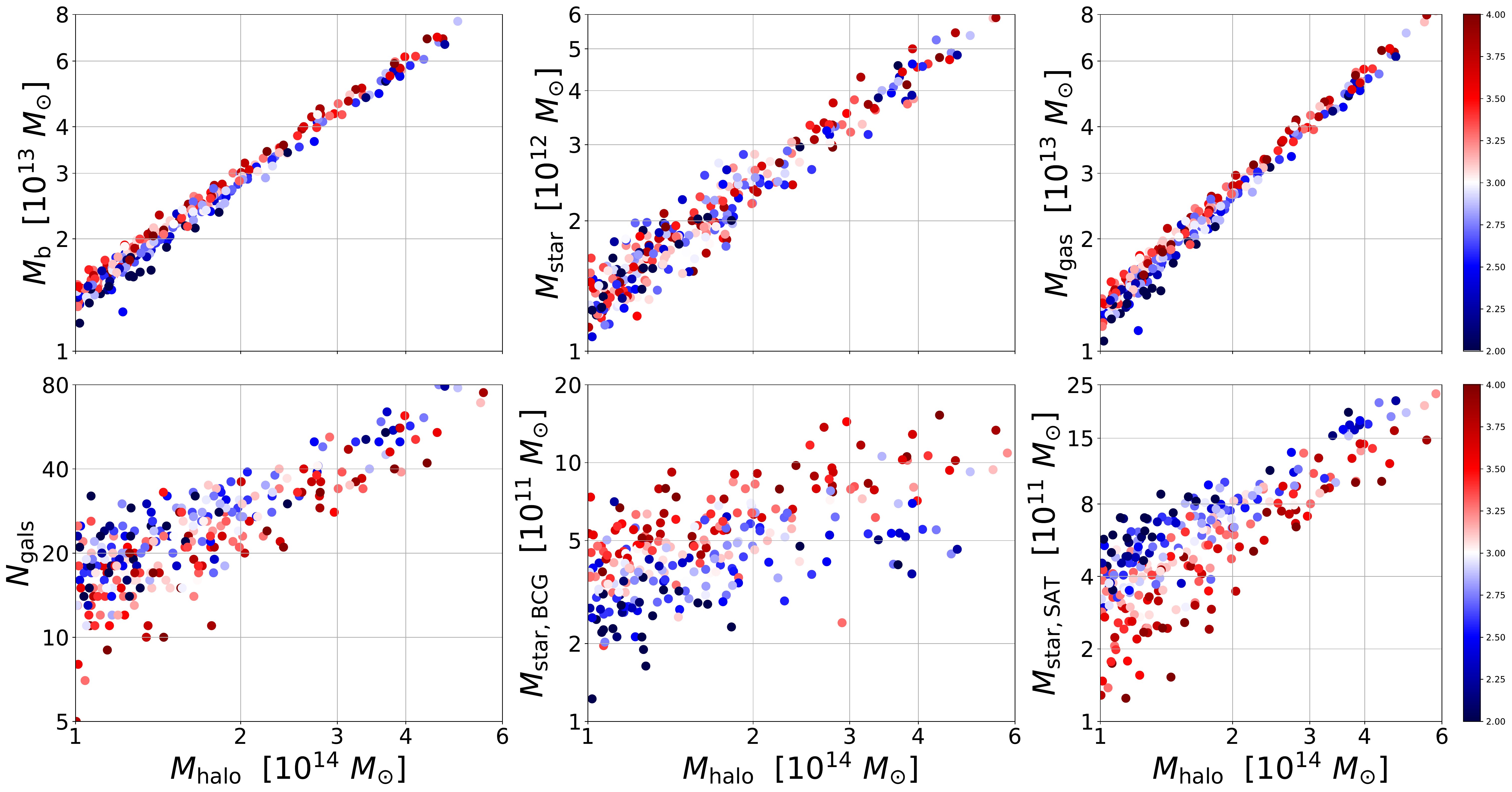}
     \caption{Halo properties vs. halo mass. Color coded according to M14 -- the magnitude difference between the brightest and fourth brightest galaxy members. Red (blue) corresponds to larger (smaller) magnitude gap. Halo properties are spherically integrated quantities within $R_{200}$ of the host halo as listed in Table \ref{tab:halo-quantities}. These results show that there is a magnitude gap stratification in halo properties at fixed halo mass. Thus, including the magnitude gap in halo property--halo mass relation reduces the intrinsic scatter. For visualization purpose, we only show a subset of halos with mass $10^{14} < M_{\rm halo}\,\, [M_{\odot}] < 6 \times 10^{14}$.} 
     \label{fig:Prop-vs-Mh-M14}
\end{figure*}

Figure \ref{fig:Prop-vs-Mh-M14} illustrates halo properties as a function of halo mass for a subset of our sample. The data points are color coded according to M14. Red (blue) points are halos with high (low) magnitude gap or equivalently early (late) forming halos. From this figure, it is evident that the majority of halo properties are correlated with the magnitude gap at fixed halo mass. We thus expect that conditioning on the magnitude gap in the HPHM relation reduces the scatter.

Table \ref{tab:sim-results} summarizes the posterior of the regression model described in Section \ref{sec:regression}. We regress halo properties against log halo mass and a secondary explanatory variable -- i.e., M12, M14, or BCG stellar mass. $\beta$ is the slope of the secondary explanatory variable. Taking M14 as a proxy of halo age, a HPHM with a positive(negative) M14 slope implies that the halo property at fixed halo mass is (anti-)correlated with the age of halo. As expected, the scatter about HPHM reduces for the correlated properties with M14. These results suggest a strong correlation between all halo quantities, except total stellar mass, and M14 at fixed halo mass.

It is previously suggested that M14 is a better indicator of the formation time than M12 \citep{Dariush:2010}. Our results show that M14 is more efficient in reducing the scatter for halo properties like the BCG stellar mass and $N_{\rm gals}$. This finding implicitly suggests that M14 is a better proxy of formation time. For the rest of this paper, we only report the conditional statistics using M14. Our conclusions and results qualitatively remain the same if instead we use M12.

\begin{table*}
\begin{tabular}{|l|l|l|l|l|}
\hline
Response property $s_p$ & $\alpha$        & $\beta$          & $\sigma$          & Secondary explanatory property $X_2$  \\ \hline
\hline
$\ln(M_{\rm b})$   & $1.033 \pm 0.006$ & $ 0.021 \pm 0.002$ & $0.050 \pm 0.002$ & M14                   \\ \hline
$\ln(M_{\rm b})$   & $1.037 \pm 0.006$ & $ 0.015 \pm 0.002$ & $0.051 \pm 0.002$ & M12                   \\ \hline
$\ln(M_{\rm b})$   & $0.994 \pm 0.008$ & $ 0.076 \pm 0.009$ & $0.052 \pm 0.002$ & $\ln(M_{\rm star, BCG})$   \\ \hline
$\ln(M_{\rm b})$   & $1.042 \pm 0.007$ & -                  & $0.058 \pm 0.003$ & -                     \\ \hline
\hline
$\ln(M_{\rm gas})$   & $1.045 \pm 0.007$ & $0.023 \pm 0.003$ & $0.057 \pm 0.002$ & M14                  \\ \hline
$\ln(M_{\rm gas})$   & $1.050 \pm 0.007$ & $0.017 \pm 0.002$ & $0.058 \pm 0.003$ & M12                  \\ \hline
$\ln(M_{\rm gas})$   & $1.008 \pm 0.010$ & $0.075 \pm 0.011$ & $0.060 \pm 0.003$ & $\ln(M_{\rm star, BCG})$  \\ \hline
$\ln(M_{\rm gas})$   & $1.056 \pm 0.008$ & -                 & $0.065 \pm 0.003$ & -                    \\ \hline
\hline
$\ln(M_{\rm star})$  & $0.896 \pm 0.011$ & $-0.001 \pm 0.004$ & $0.098 \pm 0.004$ & M14                  \\ \hline
$\ln(M_{\rm star})$  & $0.897 \pm 0.011$ & $-0.002 \pm 0.003$ & $0.098 \pm 0.004$ & M12                  \\ \hline
$\ln(M_{\rm star})$  & $0.845 \pm 0.015$ & $ 0.080 \pm 0.017$ & $0.094 \pm 0.004$ & $\ln(M_{\rm star, BCG})$  \\ \hline
$\ln(M_{\rm star})$  & $0.896 \pm 0.011$ & -                  & $0.098 \pm 0.004$ & -                    \\ \hline
\hline
$\ln(M_{\rm star, BCG})$  & $0.558 \pm 0.028$ & $0.167 \pm 0.010$ & $0.240 \pm 0.011$ & M14              \\ \hline
$\ln(M_{\rm star, BCG})$  & $0.595 \pm 0.032$ & $0.104 \pm 0.009$ & $0.275 \pm 0.012$ & M12              \\ \hline
$\ln(M_{\rm star, BCG})$  & $0.636 \pm 0.037$ & -                 & $0.333 \pm 0.015$ & -                \\ \hline
\hline
$\ln(M_{\rm star, SAT})$  & $1.156 \pm 0.027$ & $-0.176 \pm 0.010$ & $0.223 \pm 0.010$ & M14                \\ \hline
$\ln(M_{\rm star, SAT})$  & $1.121 \pm 0.029$ & $-0.118 \pm 0.009$ & $0.251 \pm 0.011$ & M12                \\ \hline
$\ln(M_{\rm star, SAT})$  & $1.446 \pm 0.043$ & $-0.584 \pm 0.049$ & $0.267 \pm 0.012$ & $\ln(M_{\rm star, BCG})$ \\ \hline
$\ln(M_{\rm star, SAT})$  & $1.075 \pm 0.037$ & -                  & $0.330 \pm 0.014$ & -                   \\ \hline
\hline
$\ln(N_{\rm gals})$  & $0.949 \pm 0.027$ & $-0.078 \pm 0.010$ & $0.227 \pm 0.010$ & M14                  \\ \hline
$\ln(N_{\rm gals})$  & $0.927 \pm 0.028$ & $-0.036 \pm 0.008$ & $0.243 \pm 0.010$ & M12                  \\ \hline
$\ln(N_{\rm gals})$  & $1.108 \pm 0.037$ & $-0.306 \pm 0.041$ & $0.230 \pm 0.010$ & $\ln(M_{\rm star, BCG})$  \\ \hline
$\ln(N_{\rm gals})$  & $0.912 \pm 0.028$ & -                  & $0.251 \pm 0.011$ & -                    \\ \hline
\end{tabular} 
\caption{The impact of secondary explanatory variable on reducing the scatter about the halo property--halo mass relation. We employ a linear log-normal model with a mean relation defined with $\langle s_{p} \, | \, \ln M_{\rm halo}, X_2 \rangle \ = \ \pi  + \alpha \, \ln M_{\rm halo} + \beta \, X_2$. $\alpha$ is slope of natural log halo mass, $\beta$ is slope of the secondary explanatory variable, and $\sigma$ is the conditional scatter of the natural log halo property at fixed halo mass and secondary explanatory variable.} \label{tab:sim-results}
\end{table*}

\subsection{Impact of formation time on total hot gas and stellar mass} \label{sec:fb-M14}

\begin{figure*}
     \centering
     \includegraphics[width=0.98\textwidth]{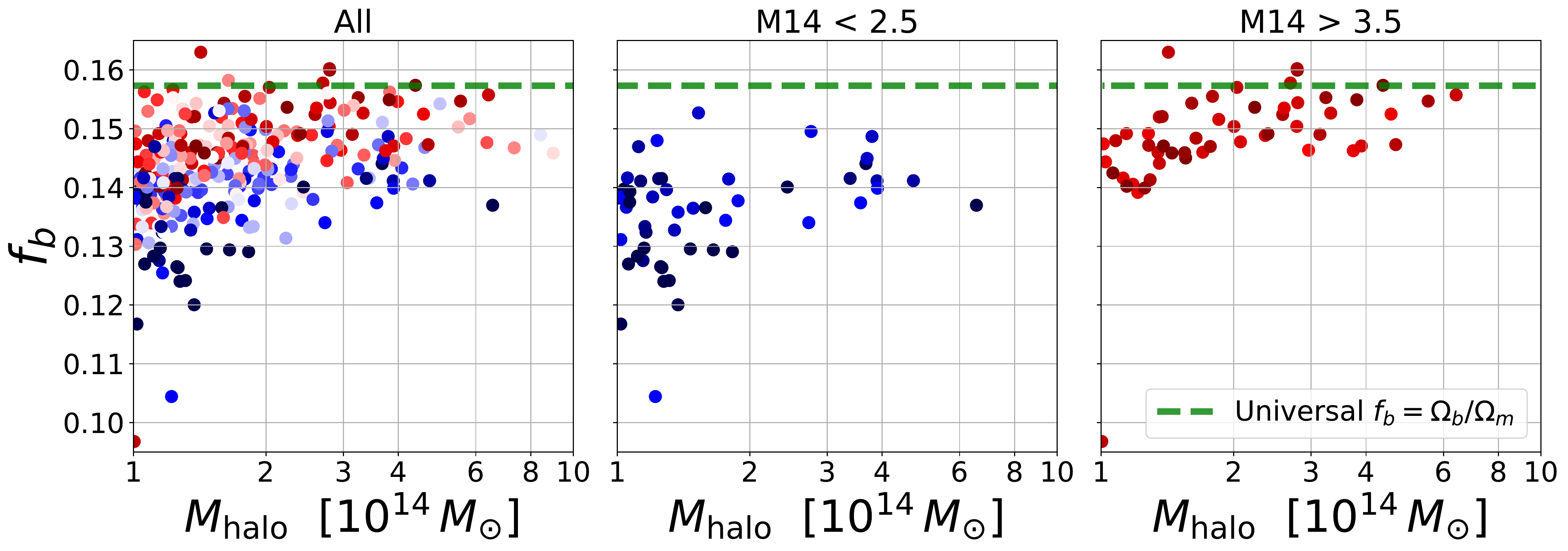}
     \caption{Baryon fraction of massive halos color coded with magnitude gap. The green dashed line shows the universal mean baryon fraction of the IllustrisTNG universe. {\bf Left Panel:} All Halos. {\bf Middle Panel:} Subset of halos with M14 smaller than 2.5. {\bf  Right Panel:} Subset of halos with M14 above 3.5. The magnitude gap color code is the same as Figure \ref{fig:Prop-vs-Mh-M14}.}
     \label{fig:fb-vs-Mh-M14}
\end{figure*}

The primary quantity that determines how much of the gas turns into stars is the amount of available gas mass. The star-formation efficiency regulates how much of the gas turn into stars. There are several competing effects control star-formation efficiency and the final stellar mass of the system. The supper massive black hole (SMBH) at the center of BCG can boost star formation by helping stir up the interstellar medium, but it can also suppress star formation in the galaxy by expelling hot gas. The SMBH and Supernovae (SN) feedback and stellar wind expel gas particles from shallow potential wells, dark matter halos of small mass \citep{Wechsler:2018}, while massive halos can retain their gas because of their deep potential wells. The correlation between hot gas mass and M14 can be explained with the latter effect. Figure \ref{fig:fb-vs-Mh-M14} shows that the baryon fraction, $M_{b} / M_{\rm halo}$, of early formed halos (the right panel) almost follows the universal baryon fraction value -- the green dashed line --  while the baryon fraction of the late formed halos (middle panel) is significantly smaller than the universal value. Our results suggest that halos formed earlier create deeper potential wells \citep{Wechsler:2002} thereby are more efficient in retaining their baryonic content, while halos formed later expel some of their baryonic content through SMBH feedback. While this likely interpretation can be drawn from our results, this is a potential explanation, and we do have no direct evidence for this over other potential causes.

The majority of the baryonic mass in halos is in form of hot gas, which cools down and forms stars. The anti-correlation between hot gas and the formation time, as implied by the positive slope of hot gas mass and M14 from Table \ref{tab:sim-results}, is expected due to the anti-correlation between total baryonic content and the formation time. The lack of correlation between total stellar mass and M14, with $\beta \approx 0$, can be explained with two competing effects: age effect and gas reservoir availability effect. Early formed halos are inefficient in forming stars, hence smaller stellar mass, but are more gas-rich to turn into stars, hence larger stellar mass. These two competing effects cancel one another, thus the correlation between formation time and total stellar mass vanishes. Another potential explanation is that the intrinsic scatter in the total stellar mass at fixed halo mass is large enough that washes out the correlation between M14 and total stellar mass. We emphasize that our data is insufficient to rule out other potential alternative interpretations of this lack of correlation. Future observational data and more comprehensive analysis of simulations outputs can help us to distinguish between these alternative hypotheses.

The mass fraction of hot gas measured within a characteristic radius of a halo at redshift $z$ can be written as
\begin{equation}
    f_{\rm gas}(z) = Y \frac{\Omega_b}{\Omega_m}\,,
\end{equation}
where $Y$ accounts for deviation from the universal baryon fraction. For the most massive systems, $Y$ is expected to be independent of the halo mass and redshift \citep{Mantz:2014fb}. Gas mass and total mass measurements from X-ray observations respond to distance differently, $\propto d^{5/2}$ and $\propto d$, respectively. These distance-dependent measurements imply that $f_{\rm gas}(z) \propto d^{3/2}(z)$. This feature can be exploited to construct a distance estimator \citep{Allen:2011review, Mantz:2014fb}. A subset of clusters with smaller $f_{\rm gas}$ intrinsic scatter can improve cosmological parameter estimation using this method. Our findings suggest that a subset of halos with a larger magnitude gap has lower scatter and reflects the mean baryonic content of the Universe more precisely. Therefore, such a subset is better suited for a cosmological parameter estimation using this method.

\subsection{Comparing with the literature}

The anti-correlation between formation time and the number of satellite galaxies is a natural consequence of hierarchical formation; the most massive galaxies inside early formed halos has more time to merge with the central galaxy by losing energy via dynamical friction, hence they have larger magnitude gap and smaller richness compare to the mean of a population of same mass halos. Examining the simulations suggested that there is an anti-correlation between formation time and the number of sub-halos at fixed halo mass \citep[{\sl e.g.},][]{Wu:2013, Jiang:2017}. The robustness of these results, however, remains to be shown, as numerical resolution is shown to affect the scatter in $P(N_{\rm gals} \, |\, M)$ \citep[][]{vandenBosch:2018, Anbajagane:2020} and might induce spurious correlation with the formation time. Observationally, \citet{Hearin:2013} established evidence of such an anti-correlation. Specifically, they found that at fixed velocity dispersion groups with small magnitude gaps are richer than groups with a larger magnitude gap. If we take velocity dispersion as a proxy for the halo mass, then their finding is qualitatively in agreement with the TNG300 expectations and the hierarchical formation scenario.

\citet{Bradshaw:2019} analyzed sources of scatter in the stellar content of halos derived from the UniverseMachine simulation \citep{Behroozi:2019}. In agreement with our findings, they find that the scatter in total stellar mass, in oppose to the BCG stellar mass, is uncorrelated with the age of halo \citep[see Figure 4 in][]{Bradshaw:2019}. The age--BCG stellar mass correlation implicitly suggests that the more massive BCGs, at fixed halo mass, grown through major mergers and lends support to the bottom-up hierarchical formation scenario. \citet{Bradshaw:2019} argue that total stellar mass has no strong environment or secondary parameter dependence, and can, therefore, be approximated by a model that only accounts for halo mass, such as the conditional luminosity function. As discussed in Section \ref{sec:fb-M14}, we speculate that there are two competing effects, age effect, and available gas reservoir effect, that cancel the effects of one another. This hypothesis suggests that the scatter in total stellar mass at fixed halo mass can be reduced by incorporating these competing parameters simultaneously.

Observational studies have established a correlation between BCG stellar mass and magnitude gap at fixed inferred halo mass \citet{GoldenMarx:2018, Golden-Marx:2019}. \citet{GoldenMarx:2018} employed a subset of optically-selected SDSS clusters with a caustic mass measurement to study the relation between magnitude gap and BCG stellar mass. Their results show a cluster with a larger(smaller) magnitude gap corresponds to a more(less) massive BCG with respect to the mean relation. They have shown that incorporating this observable reduces the scatter in cluster mass--BCG stellar mass relation from $\sim 0.36 \pm 0.05$ to $0.19 \pm 0.06$, which is both qualitatively and quantitatively in agreement with our results\footnote{\citet{GoldenMarx:2018} report the scatter in $\log_{10}$-based while we report the scatter in $\ln$-based quantities. To compare with our results, we convert their numbers into $\ln$-based quantities.}. The reader should bear in mind that with a larger sample and smaller statistical uncertainties, the systematics such as the projection effect need to be understood better before trying to compare the observational findings with simulations results directly.

With a sample of 41 X-ray bright clusters, \citet{Mulroy:2019} studied the mass--observable scaling relation for a broad range of optical, X-ray, and millimeter wavelengths observables. They plot cluster observable residuals about the mean relation as a function of magnitude gap \citet[see Figure C1 of][]{Mulroy:2019}. A visual inspection of their data suggests that BCG $K$-band luminosity and $M_{\rm gas}$ at fixed weak-lensing mass are positively correlated with M12, while optical-richness residuals are anti-correlated with M12. Even-though, the authors did not quantify these correlations, their data suggest that the trend is in qualitative agreement with our findings in Table \ref{tab:sim-results}.

\subsection{Log-normality of residuals} \label{sec:log-normal}

The log-normal model is a core assumption of many cluster cosmology parameter inference and scaling relation analysis \citep[{\sl e.g.},][]{Pratt:2009, Vikhlinin:2009scaling, Mantz:2010, Mulroy:2019, Farahi:2019}. \citet{Evrard:2014} employed this model to construct a powerful mathematical framework for conditional statistics of observables given the log-normal assumption. Their mathematical model is agnostic about the physics, thus assessing the relative accuracy of their assumptions is of particular importance. In this section, we study the log-normality of conditional halo properties -- with and without conditioning on M14.

We express the conditional normalized residuals of a halo property with $\widehat{\delta} s_{a}$, where 
\begin{equation} \label{eq:normalized-residuals}
    \widehat{\delta} s_{a,i} \equiv \frac{\delta s_{a,i}}{\sigma_a } 
    = \frac{ s_{a,i} - \langle s_{a,i} | M_i,X_{2,i} \rangle  } {\sigma_a }  
\end{equation}
$\sigma_a$ is a log-normal scatter and $\langle s_{a,i}\, |\, M_i,X_{2,i} \rangle$ is the mean halo property at given mass and the secondary explanatory variable, defined in Section \ref{sec:regression}, and $s_a$ is a logarithmic halo property. Similar to \citet{Shaw:2010}, we formulate the non-Gaussianiny of conditional distributions in terms of an Edgeworth series expansion, 
\begin{equation}
    P(s_a \, | \, M, X_2) \approx G(\cdot) - \frac{\gamma}{6}\frac{{\rm d}^3G}{{\rm d}x^3} ,
\end{equation}
\noindent where the skewness, $\gamma$, is defined as, 
\begin{equation}
\gamma = \left\langle\left( \frac{s_a - \langle s_a \rangle}{\sigma}  \right)^3 \right\rangle\,,
\end{equation}
\noindent and $G(\cdot)$ is a Gaussian distribution conditioned on halo mass and the secondary explanatory variable.

\begin{figure*}
    \centering
    \includegraphics[width=0.98\textwidth]{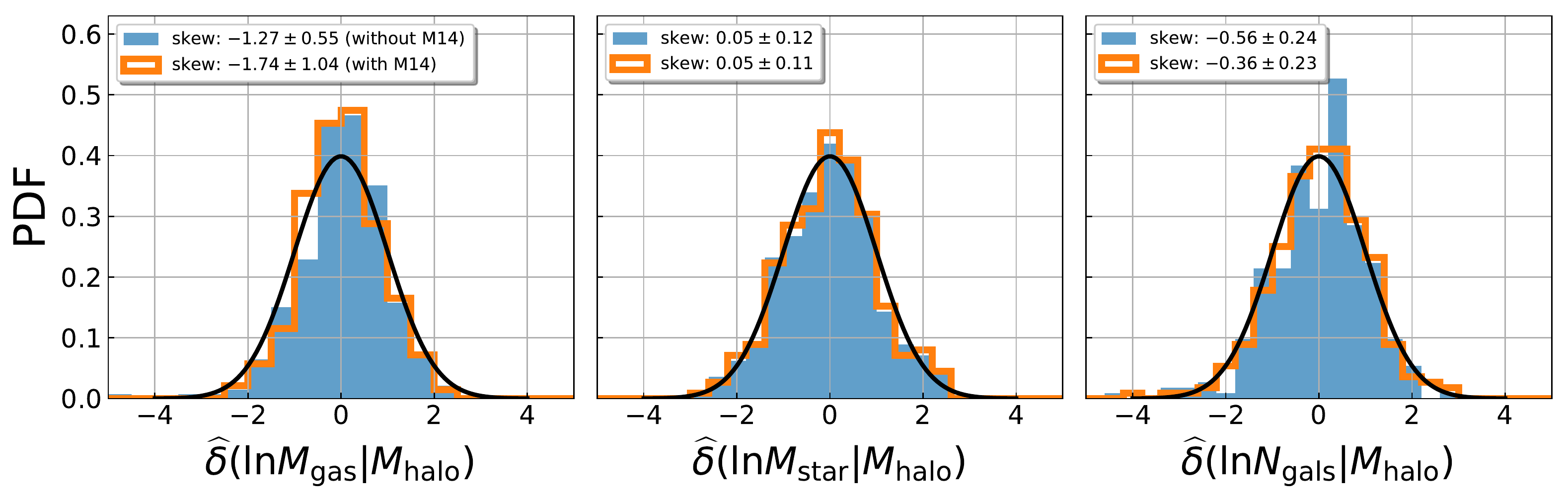}
    \caption{Distribution of conditional normalized residual, Equation (\ref{eq:normalized-residuals}), for gas mass (left), stellar mass (middle), and richness (right). The orange histograms show distribution of normalized residuals conditioned on halo mass and M14 while the blue histograms are normalized residuals conditioned only on halo mass.}
    \label{fig:residuals-shape}
\end{figure*}

Figure \ref{fig:residuals-shape} shows the distribution of $\widehat{\delta}M_{\rm star}$, $\widehat{\delta}M_{\rm gas}$, and $\widehat{\delta}N_{\rm gals}$. Each orange histogram shows normalized residuals of a halo property conditioned on halo mass and M14, while the corresponding blue histogram shows same statistics without conditioning on M14. The mean, standard deviation, and skewness of residuals are specified in the legend of each figure. The error bars are computed with generating 1,000 bootstrap realizations of residuals.  

Assuming a power-law mass function $[{\rm d}n/{\rm d} \ln M] \propto M^{-\alpha}$, cluster number count can be expressed as 
\begin{equation}
    \frac{{\rm d} n}{{\rm d} \ln M} \approx \left( \frac{{\rm d} n}{{\rm d} \ln M} \right)_0 \times (1 + \frac{\alpha^3 \sigma^3}{6} \gamma + \cdots )
\end{equation}
where $\sigma$ is the natural logarithmic prediction scatter, and $\left( \frac{{\rm d} n}{{\rm d} \ln M} \right)_0$ is the number density assuming a log-normal model \citep[see, equation (156) of][]{Weinberg:2013}. We assume a power law of slope $\alpha = 2$ for the halo mass function \citep[see, Table 1 of][]{Evrard:2014}. Thus, achieving sub-percent level systematic uncertainty in cluster number counts under a log-normal approximation with a mass proxy having $20\% \,\,(10\%)$ scatter requires roughly $\gamma < 1\,\, (8)$ \citep{Shaw:2010, Weinberg:2013}. 

We note that the estimated skewness is smaller than what is expected for sub-percent systematics requirement, and conditioning on M14 does not increase the non-Gaussianity of residuals. To achieve sub-percent systematics, lower the scatter a higher value of skewness can be tolerated. Reducing the scatter decreases the impact of non-Gaussianity in a cluster number count analysis with a factor of $\sigma^3$. This implies that including the secondary explanatory variable, like M14, has the advantage of reducing the impact of non-log-normal systematics.

\citet{Anbajagane:2020} study the conditional statistics of stellar properties of cluster and group size halos realized by three different cosmological hydrodynamics simulations --- IllustrisTNG, BAHAMAS+MACSIS, and Magneticum. They illustrate that the mass-conditioned likelihoods of the three simulations follow a common skewed log-normal shape, i.e. $P(N_{\rm gals} \,|\, M_{\rm halo}, z)$ being skew log-normal, with skewness decreasing with mass. Even though their definition of satellite galaxy slightly differs from our definition, they find a similar skewness for a subset of halos with $M_{200} > 10^{14}\, M_{\odot}$. They also illustrate that the low $N_{\rm gals}$ population is driven by early forming halos. This finding supports the hierarchical growth scenario and is in agreement with our results.

When modeling observational data, the form of the conditional statistics of measured quantities may differ from a log-normal form, for example due to projection effects \citep[{\sl e.g.}, ][]{Cohn:2007, Erickson:2011, Costanzi:2018projection}. Analysis of such data using a log-normal assumption in the likelihood adds a systematic bias in mass--observable relation that leads to bias in a cosmological parameter estimation. These additional uncertainties are strongly dependent on survey characteristics and data reduction pipelines, thus must be modeled for each survey independently. Investigation of these additional uncertainties is beyond the scope of this work.

\subsection{Halo Property Pair Correlation}
\label{sec:halo-prop-corr}

Halo property pair correlation is another observable which carries information about astrophysics \citep{Farahi:2019anti}. This quantity is the correlation in the scatter of a pair of halo properties at fixed halo mass. The magnitude and sign of this correlation are sensitive to sub-grid models and how the baryons evolve within the potential well of the host dark matter halo \citep{Stanek:2010, Wu:2015, Truong:2018, Farahi:2018bahamas}. A precise and accurate estimation of this quantity will allow us to improve our understanding of sub-grid physics.

\begin{figure*}
    \centering
    \includegraphics[width=0.98\textwidth]{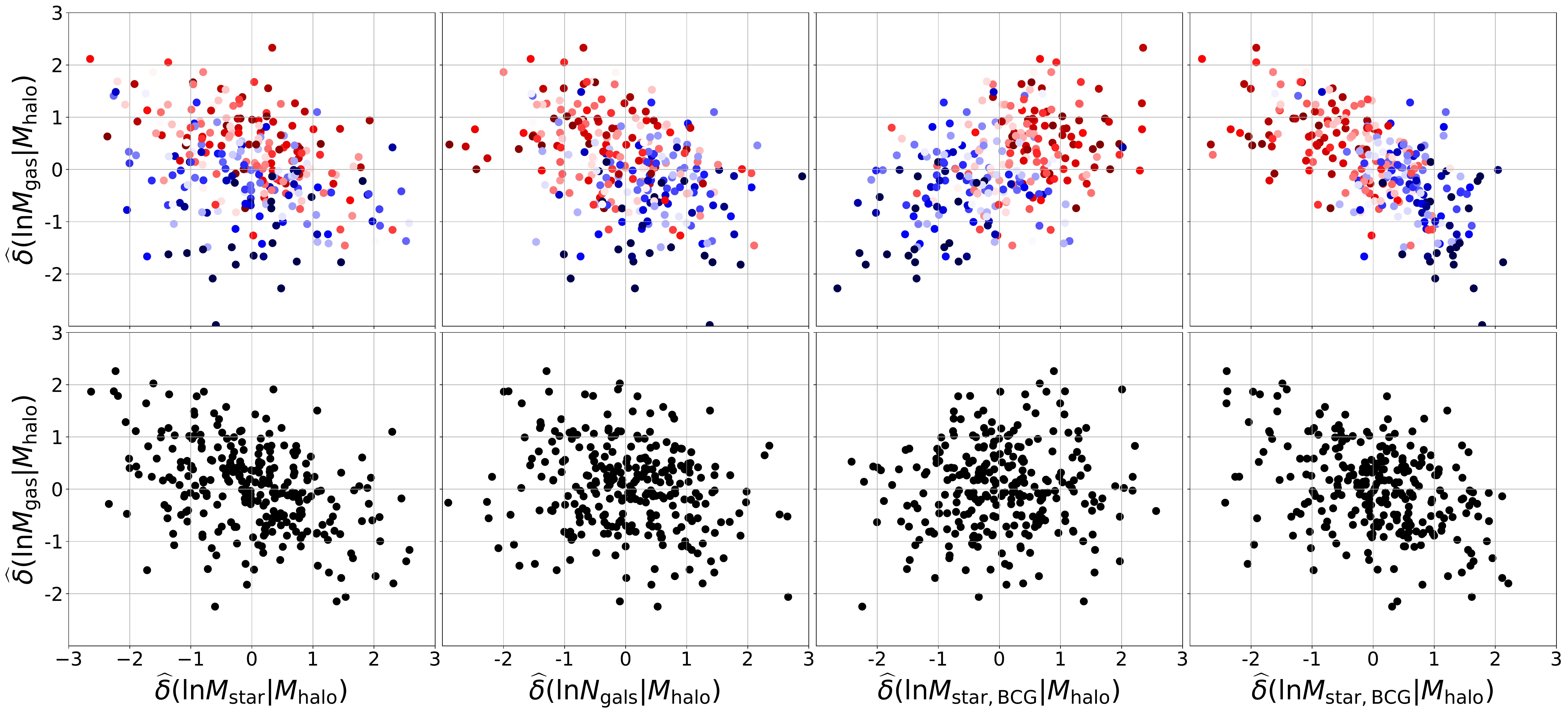}
    \caption{Conditional normalized residual of gas mass, Equation (\ref{eq:normalized-residuals}), vs. conditional normalized residual of other halo properties. The top panels are normalized residuals conditioned on halo mass only, and the bottom panels are normalized residuals conditioned on halo mass mad M14. The residuals on the top panels are colored by M14. M14 color code is the same as Figure \ref{fig:Prop-vs-Mh-M14}.}
    \label{fig:residuals-corr-Mgas}
\end{figure*}

\begin{figure*}
    \centering
    \includegraphics[width=0.98\textwidth]{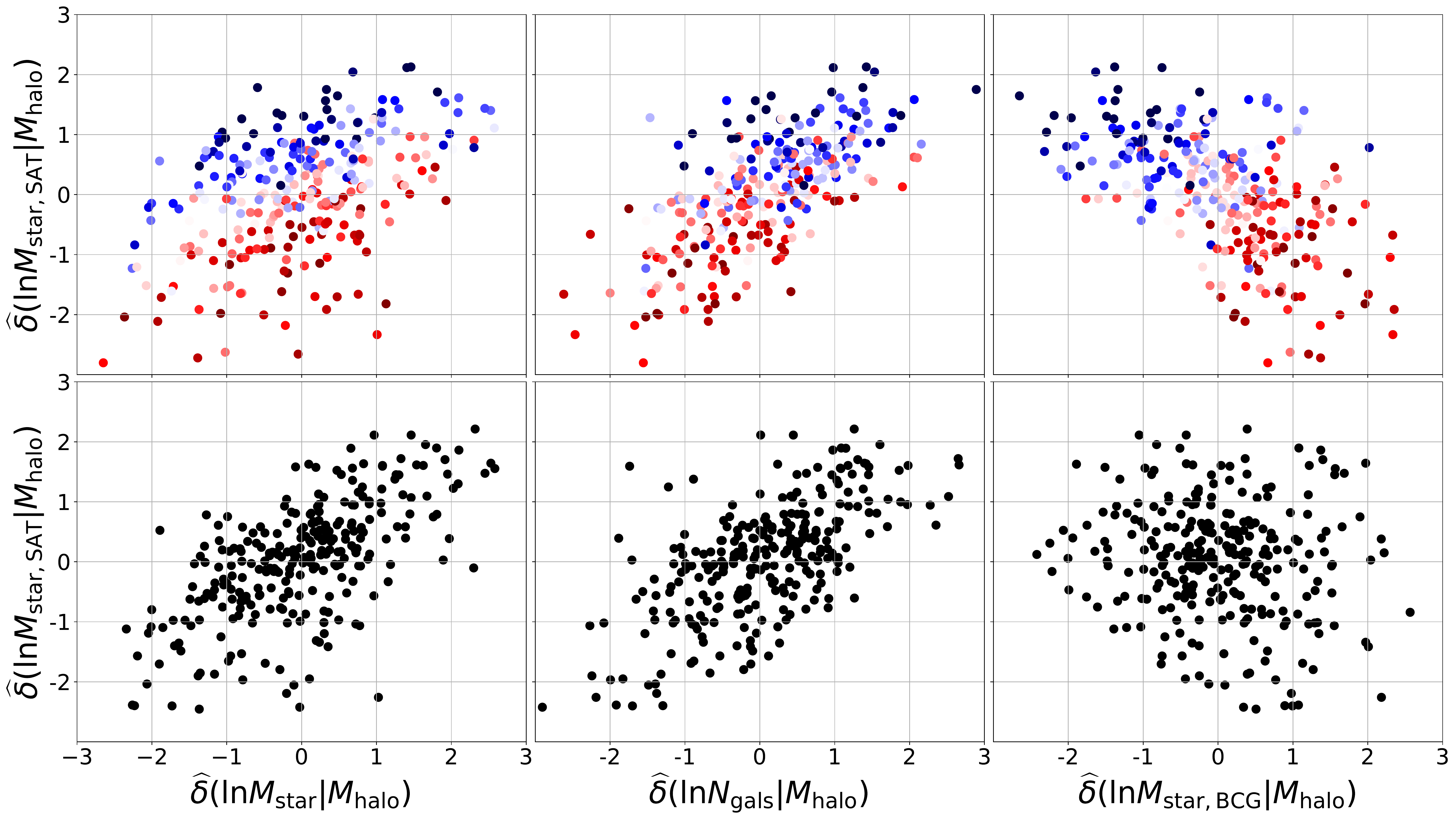}
    \caption{Same as Figure \ref{fig:residuals-corr-Mgas}, except the y-axis shows conditional normalized residual for stellar mass of the satellite galaxies. }
    \label{fig:residuals-corr-Msat}
\end{figure*}

The lower diagonal of Table \ref{tab:property-pair-correlation} presents halo property pair correlations, and the diagonal elements are the scatter of halo properties. The top table presents halo property covariance conditioned on halo mass, while the bottom table shows halo property covariance conditioned on halo mass and M14. Conditioning halo properties on M14 and halo mass allows us to eliminate the impact of formation time and study the conditional statistics of similar mass and age halos. Figure \ref{fig:residuals-corr-Mgas} and Figure \ref{fig:residuals-corr-Msat} illustrate the correlation structure of normalized residuals for a subset of halo properties. In the top panels, the normalized residuals are only conditioned on halo mass and are colored by M14; and in the bottom panels, the normalized residuals are conditioned on halo mass and M14.

From Figure \ref{fig:residuals-corr-Mgas} and Figure \ref{fig:residuals-corr-Msat} it is evident that when residuals are correlated with M14 along the correlation axis conditioning on M14 reduces the correlation coefficient; when residuals are uncorrelated with M14 conditioning on M14 does not change the correlation coefficient; and when residuals are correlated with M14 perpendicular to the correlation axis conditioning on M14 increases the correlation coefficient. We speculate that physically there are two competing scenarios that determine how the correlation coefficient changes when halo properties are conditioned on the formation time: one reduces the correlation, and one makes the correlation stronger.  If halo properties are not conditioned on the formation time, then properties that are strongly correlated with age are expected to become correlated. This correlation vanishes when halo properties conditioned on the formation time. In the second scenario, we assume two halo properties of similar age are intrinsically correlated. By marginalizing over the formation time, halos of different ages get mixed. This mixing induces additional scatter; therefore, the correlation coefficient becomes weaker. In the second scenario, conditioning on the formation time in a regression analysis allows us to recover the intrinsic correlation. These two competing scenarios determine if the (anti-)correlation coefficient should become stronger or weaker.

A key quantity in Table \ref{tab:property-pair-correlation} is the anti-correlation between the gas and stellar content of clusters. This anti-correlation has been noticed in hydrodynamical simulations \citep{Wu:2015, Farahi:2018bahamas}, and is recently measured via observational data \citep{Farahi:2019anti}. This anti-correlation can be a result of halos behaving like closed baryon boxes. If the baryonic content of dark matter halos is retained within the boundary of halo, such anti-correlation is naturally expected. Our results provide yet another measurement of this anti-correlation from an independent hydrodynamical simulation. Halos of similar mass and age probe clusters that are formed in a similar environment. Conditioning on M14 eliminates this source -- halo age -- of scatter. Thus, we speculate that the anti-correlation should become stronger. The statistical uncertainty of our sample is quite large with respect to change in the value of the correlation coefficient. So statistically speaking, we cannot rule out a scenario in which the anti-correlation remains the same or even becomes weaker.

Conditioning on M14 reduces the (anti-)correlation of halo properties which are mainly driven by the formation time. The correlation between gas mass and BCG stellar mass at fixed halo mass is likely driven by the formation time. A younger population of halos of similar mass have less gas available to them, and their BCG is less massive than mean. Therefore, conditioning on M14 as a proxy for age should reduce this correlation. We indeed find that the correlation coefficient between hot gas mass and BCG stellar mass becomes close to zero when we condition the halo properties on M14 and halo mass. The accretion and tidal disruption of satellite galaxy stellar material onto the central galaxy is a mechanism that can induce an anti-correlation between $M_{\rm star, SAT}$ and $M_{\rm star, BCG}$. For an older population of halos of the same mass, satellite galaxies have more time to accrete into the central galaxy, which results in fewer satellite galaxies and a more massive BCG. This explanation implies that conditioning BCG stellar mass and $N_{\rm gals}$ on M14, in addition to halo mass, should wash out this anti-correlation, similarly for the anti-correlation between $M_{\rm star, SAT}$ and $M_{\rm star, BCG}$.

We notice that $M_{\rm star, SAT}$ and $M_{\rm star}$ become strongly correlated after conditioning on halo mass and M14. This finding indicates that the stellar mass of satellite galaxies of halos with similar mass and age is a strong predictor of the total stellar mass of the host halo. The reduce in the correlation coefficient by marginalizing over M14 can be explained with an additional scatter due to the mixing halos of different age. This additional scatter reduces the correlation between these two intrinsically correlated quantities while one is uncorrelated with age, here $M_{\rm star}$, and the other one is correlated with age. Reducing this source of scatter makes $M_{\rm star, SAT}$ and $M_{\rm star}$ more correlated (see the left panels of Figure \ref{fig:residuals-corr-Msat}). In Appendix \ref{app:covariance-splitting}, we provide a mathematical model that can quantify this effect.

\begin{table*}
\caption{The impact of conditioning on M14 on the property pair covariance of halos of the same mass. {\bf Lower Triangle}:  The property pair correlation coefficients. The uncertainties are computed using 1000 bootstrap realizations. Cell color encodes the magnitude and sign of the correlation coefficient, with red (blue) showing positive (negative) values. {\bf Diagonal (grey cells)}: Intrinsic scatter of halo properties.  The uncertainties are computed using 1000 bootstrap realizations. } \label{tab:property-pair-correlation}
\begin{tabular}{|l|l|l|l|l|l|}
\hline
\multicolumn{6}{|l|}{\textbf{(a) Conditioned on halo mass.}} \\
\rowcolor[HTML]{A9A9A9} \cellcolor[HTML]{333333}   & $M_{\rm gas}$  & $M_{\rm star}$  & $N_{\rm gals}$  & $M_{\rm *, BCG}$  & $M_{\rm *, SAT}$   \\ \hline
\cellcolor[HTML]{A9A9A9} $M_{\rm gas}$   & \cellcolor[HTML]{CCCCCD}  $ 0.028 \pm 0.002 $    & \cellcolor[HTML]{333333}   & \cellcolor[HTML]{333333}   & \cellcolor[HTML]{333333}   & \cellcolor[HTML]{333333}   \\ \hline
\cellcolor[HTML]{A9A9A9} $M_{\rm star}$   & \cellcolor[rgb]{0.745, 0.822, 1.000}  $ -0.255 \pm 0.074 $
  & \cellcolor[HTML]{CCCCCD}  $ 0.042 \pm 0.002 $    & \cellcolor[HTML]{333333}   & \cellcolor[HTML]{333333}   & \cellcolor[HTML]{333333}   \\ \hline
\cellcolor[HTML]{A9A9A9} $N_{\rm gals}$   & \cellcolor[rgb]{0.753, 0.827, 1.000}  $ -0.247 \pm 0.118 $
  & \cellcolor[rgb]{1.000, 0.763, 0.661}  $ 0.339 \pm 0.052 $
  & \cellcolor[HTML]{CCCCCD}  $ 0.108 \pm 0.006 $    & \cellcolor[HTML]{333333}   & \cellcolor[HTML]{333333}   \\ \hline
\cellcolor[HTML]{A9A9A9} $M_{\rm *, BCG}$   & \cellcolor[rgb]{1.000, 0.728, 0.611}  $ 0.389 \pm 0.073 $
  & \cellcolor[rgb]{1.000, 0.809, 0.727}  $ 0.273 \pm 0.052 $
  & \cellcolor[rgb]{0.595, 0.716, 1.000}  $ -0.405 \pm 0.048 $
  & \cellcolor[HTML]{CCCCCD}  $ 0.143 \pm 0.006 $    & \cellcolor[HTML]{333333}   \\ \hline
\cellcolor[HTML]{A9A9A9} $M_{\rm *, SAT}$   & \cellcolor[rgb]{0.473, 0.631, 1.000}  $ -0.527 \pm 0.105 $
  & \cellcolor[rgb]{1.000, 0.691, 0.558}  $ 0.442 \pm 0.048 $
  & \cellcolor[rgb]{1.000, 0.495, 0.278}  $ 0.722 \pm 0.033 $
  & \cellcolor[rgb]{0.413, 0.589, 1.000}  $ -0.587 \pm 0.035 $
  & \cellcolor[HTML]{CCCCCD}  $ 0.142 \pm 0.007 $    \\ \hline
 \hline
\multicolumn{6}{|l|}{\textbf{(b) Conditioned on halo mass and M14.}} \\
\rowcolor[HTML]{A9A9A9} \cellcolor[HTML]{333333}   & $M_{\rm gas}$  & $M_{\rm star}$  & $N_{\rm gals}$  & $M_{\rm *, BCG}$  & $M_{\rm *, SAT}$   \\ \hline
\cellcolor[HTML]{A9A9A9} $M_{\rm gas}$   & \cellcolor[HTML]{CCCCCD}  $ 0.024 \pm 0.003 $    & \cellcolor[HTML]{333333}   & \cellcolor[HTML]{333333}   & \cellcolor[HTML]{333333}   & \cellcolor[HTML]{333333}   \\ \hline
\cellcolor[HTML]{A9A9A9} $M_{\rm star}$   & \cellcolor[rgb]{0.705, 0.794, 1.000}  $ -0.295 \pm 0.089 $
  & \cellcolor[HTML]{CCCCCD}  $ 0.042 \pm 0.002 $    & \cellcolor[HTML]{333333}   & \cellcolor[HTML]{333333}   & \cellcolor[HTML]{333333}   \\ \hline
\cellcolor[HTML]{A9A9A9} $N_{\rm gals}$   & \cellcolor[rgb]{0.955, 0.968, 1.000}  $ -0.045 \pm 0.124 $
  & \cellcolor[rgb]{1.000, 0.740, 0.628}  $ 0.372 \pm 0.050 $
  & \cellcolor[HTML]{CCCCCD}  $ 0.098 \pm 0.005 $    & \cellcolor[HTML]{333333}   & \cellcolor[HTML]{333333}   \\ \hline
\cellcolor[HTML]{A9A9A9} $M_{\rm *, BCG}$   & \cellcolor[rgb]{1.000, 0.945, 0.921}  $ 0.079 \pm 0.061 $
  & \cellcolor[rgb]{1.000, 0.730, 0.615}  $ 0.385 \pm 0.052 $
  & \cellcolor[rgb]{0.832, 0.882, 1.000}  $ -0.168 \pm 0.052 $
  & \cellcolor[HTML]{CCCCCD}  $ 0.103 \pm 0.006 $    & \cellcolor[HTML]{333333}   \\ \hline
\cellcolor[HTML]{A9A9A9} $M_{\rm *, SAT}$   & \cellcolor[rgb]{0.717, 0.802, 1.000}  $ -0.283 \pm 0.106 $
  & \cellcolor[rgb]{1.000, 0.548, 0.354}  $ 0.646 \pm 0.039 $
  & \cellcolor[rgb]{1.000, 0.535, 0.335}  $ 0.665 \pm 0.039 $
  & \cellcolor[rgb]{0.843, 0.890, 1.000}  $ -0.157 \pm 0.055 $
  & \cellcolor[HTML]{CCCCCD}  $ 0.096 \pm 0.004 $    \\ \hline
\end{tabular}
\end{table*}

\section{Discussion}\label{sec:discussion}

In this section, we discuss some implications of the magnitude gap stratification, other potential secondary explanatory variables, and the key systematics in comparing with empirical data.

\subsection{Implication of formation time for the self-similar model}

Scaling relations can be scattered by a number of processes acting in different directions. Non-thermal sources of gas pressure, temperature inhomogeneity, substructures and clumps, unvirialized bulk motions, subsonic turbulence, and finally the formation time play a role. Self-similar solution can be corrected by accounting for the impact of these physical processes.  Here we study the impact of formation time on the slope of the scaling relations. We take a backward approach and ask how the slope of HPHM relation changes by conditioning on halo mass and M14.

Conditioning on M14 brings  slope of halo mass--gas mass and halo mass--total baryon mass relations closer to the self-similar expectation (see Table \ref{tab:halo-quantities}).  Conditioning on M14 and halo mass allows us to probe halos of similar formation time and assembly history. These halos are essentially formed at the same time. One potential interpretation is that the self-similar model is a better predictor of halos with similar formation time and assembly history.

Here we only study the age effect through the magnitude gap, but baryonic processes can also disrupt the self-similar relations. SMBH and SN feedback, stellar wind, or radiative cooling can remove cold, dense gas from the inner regions of low mass halos, which makes the gas mass--total mass and luminosity--total mass relations steeper and the temperature--total mass relation shallower \citep{Truong:2018}. A future direction would be to incorporate these baryonic processes into the scaling analysis and study the impact of baryonic processes and formation history simultaneously.

\subsection{Hot gas observables as secondary explanatory variable}

In this work, we study the impact of the magnitude gap on explaining the scatter about HPHM. In principle, this analysis can be extended to dynamical state and other formation time proxies. For example, the distribution and morphological properties of hot gas, which are observable through X-ray and millimeter wavelenghts, can inform us about the dynamical state and the assembly history of cluster \citep{Parekh:2015}.

The X-ray luminosity and temperature of hot gas depend on the assembly history of the clusters \citep{Hartley:2008, Fujita:2018}. Hot gas temperature increases after a major merger for a short time which is required for the shock to propagate throughout the cluster medium \citep{Fujita:2018}. Baryonic properties are sensitive to assembly history of the dark matter halo and move coherently along the scaling relation. For example, with a luminosity thresholded sample of 41 nearby clusters, \citet{Mulroy:2019} have shown that there is a strong correlation between the central gas entropy and residual of cluster observables about the mean relation. Morphological properties of the hot gas can also be employed to construct a more complex mass--observable relation to reduce the scatter \citep{Green:2019}. An interesting direction would be to check if the X-ray morphological parameters are a better indicator of the age or can more efficiently reduce the scatter.

\subsection{Improving mass estimation}

We now ask given a set of halo observables how precise the total mass of dark matter halo can be estimated. To answer this question, we employ a linear model as before with the log mass as the response variable, $\ln M_{\rm halo} \sim \mathcal{N}(\langle \ln M_{\rm halo} \, | \, \textbf{s}_p \rangle, \sigma_{m})$. We model the mean log halo mass with a weighted linear sum of baryonic observables, $\langle \ln M_{\rm halo} \, | \, \textbf{s}_p \rangle = \pi + \sum \alpha_{i} s_i$. The scatter follow a normal distribution with variance $\sigma^2_{m}$. In Table \ref{tab:reduced-mass-scatter-sims}, we report the results of this regression model for different subsets of observables. Our results suggest that a linear combination of multiple optical and X-ray observables can reduce the scatter in the estimated log halo mass. Specifically, a linear model with $M_{\rm gas}$ and $M_{\rm star}$ improves the log halo mass prediction by $\sim 14\%$ with respect to the benchmark model, when the total baryonic content of clusters is used as one parameter.

It is hardly surprising that by integrating observables into single quantity or reporting one summary statistic, like the mean or the second moment in case of velocity dispersion, we lose information about the full distribution. It has been shown that by incorporating higher-order statistics one can reduce the scatter in halo mass estimation \citep[{\sl e.g.},][]{Armitage:2019, Green:2019}.  Previous studies of the dynamical properties of cluster member galaxies have illustrated that exploiting full distribution of LOS of velocities of galaxies can improve the mass estimation by a factor of $3$ \citep{Ntampaka:2015, Ntampaka:2016, Ho:2019}. In this work, we still rely on the integrated mean quantities, but our results suggest that one can gain information about the halo mass by splitting integrated quantities into smaller pieces.

With optical-surveys, optical-richness \citep{Rykoff:2016, Farahi:2019} and total stellar mass \citep{Palmese:2019} have emerged as a potential low scatter proxy of halo mass to perform cosmological analysis \citep{SDSS:2018cosmology}. The statistical parametric models, such as the linear model employed here, are readily applicable to observational data, and the parameters can be inferred directly from the data \citep{McClintock:2019}. Our results suggest that splitting the total stellar mass into satellite and BCG components might improve the precision of halo mass estimation, which can be explained via the correlation between BCG mass and the age of halo \citep{Bradshaw:2019}. A linear combination of BCG and satellite stellar mass can maintain the age information, which reduces a source of scatter in halo mass estimation. We speculate that exploiting the full distribution of galaxy properties can further reduce scatter on halo mass estimation. This can be achieved through a machine-learning method or a more sophisticated statistical method than the one used in this work.

Machine learning methods are very efficient and powerful in combining multiple observables to build a precise mass estimator \citep{Armitage:2019, Cohn:2019, Ho:2019}. However, these methods come with their costs. These methods rely on highly non-linear models that are very hard, though not impossible \citep{Ntampaka:2019}, to interpret. They typically require a large training sample, which is hard to curate and computationally expensive to simulate. Most importantly, the trained models are susceptible to the input physics of the training sample, and estimating the measurement uncertainties are not a straightforward task. Finally, incorporating the sample selection into these models is a challenging task. More in-depth research programs are needed to address these limitations before machine learning can become a competitive analysis tool \citep{Ntampaka:2019MLinAstro}.

\begin{table*}
\begin{tabular}{|l|l|l|l|l|l|l|}
\hline
$s_1$ & $\alpha_1$ & $s_2$ & $\alpha_2$ & $s_3$ & $\alpha_3$ & $\sigma_m$     \\ \hline
\hline
$\ln M_{\rm b}$         & $0.948 \pm 0.006$ & -                       & -                 & -                       & -                  & $0.055 \pm 0.002$ \\ \hline
$\ln M_{\rm gas}$       & $0.934 \pm 0.007$ & -                       & -                 & -                       & -                  & $0.061 \pm 0.002$ \\ \hline
$\ln M_{\rm star}$      & $1.068 \pm 0.013$ & -                       & -                 & -                       & -                  & $0.107 \pm 0.005$ \\ \hline
\hline
$\ln M_{\rm gas}$       & $0.669 \pm 0.020$ & $\ln M_{\rm star}$      & $0.318 \pm 0.023$ & -                       & -                  & $0.047 \pm 0.002$ \\ \hline
$\ln M_{\rm gas}$       & $0.967 \pm 0.009$ & $\ln M_{\rm star, BCG}$ &$-0.053 \pm 0.011$ & -                       & -                  & $0.059 \pm 0.003$ \\ \hline
$\ln M_{\rm gas}$       & $0.830 \pm 0.009$ & $\ln M_{\rm star, SAT}$ & $0.107 \pm 0.008$ & -                       & -                  & $0.047 \pm 0.002$ \\ \hline
$\ln M_{\rm star, SAT}$ & $0.552 \pm 0.017$ & $\ln M_{\rm star, BCG}$ & $0.478 \pm 0.023$ & -                       & -                  & $0.165 \pm 0.007$ \\ \hline
$\ln M_{\rm star}$      & $1.038 \pm 0.028$ & $\ln N_{\rm gals}$      & $0.030 \pm 0.026$ & -                       & -                  & $0.107 \pm 0.004$ \\ \hline
\hline
$\ln M_{\rm gas}$       & $0.810 \pm 0.014$ & $\ln M_{\rm star, SAT}$ & $0.116 \pm 0.009$ & $\ln M_{\rm star, BCG}$ & $0.019 \pm 0.011$  & $0.047 \pm 0.002$ \\ \hline
\end{tabular} \caption{The statistical power of linear combination of multiple observables on predicting total halo mass. We employ a linear log-normal model, $\ln M_{\rm halo} \sim \mathcal{N}(\pi + \sum \alpha_{i} s_i, \sigma_{m})$, that map a vector of observables $\textbf{s}_p$ to log halo mass. $\sigma_m$ quantifies the intrinsic uncertainty of each model.  } \label{tab:reduced-mass-scatter-sims}
\end{table*}

\subsection{Implications for Fossil Groups} 

Fossil groups were first proposed by \citet{Ponman:1994}, when they found that an isolated galaxy was surrounded by an X-ray halo typical of a galaxy group. Since then, these systems received a lot of attention. Fossil groups are groups of galaxies associated with a large early-type galaxy dominated by an extended X-ray source of at least $L_X > 10^{42}\,{\rm erg}\,s^{-1}$. \citep{Jones:2003} proposed M12 $> 2$ in r-band magnitude as a criterion to split a cluster sample into fossil groups and non-fossil groups, later \citet{Dariush:2010} proposed M14 $>2.5$ as a more reliable indicator for this classification. In summary, the fossil systems need to have high X-ray luminosity, large magnitude gap, and a massive central galaxy.

These systems naturally emerge from our sample. At fixed halo mass, halos with a massive BCG have large gas mass and large magnitude gap (compare to the mean). Halo properties smoothly varies with the magnitude gap. This smooth transition suggests that there is no fossil/non-fossil dichotomy. Physically the lack of bright satellite galaxies is linked to the cannibalism of the central galaxy, i.e. the central galaxy is grown by major merger \citep{Corsini:2018}. \citet{Kundert:2017} argued that the observed magnitude is mostly driven by the merging of early arriving massive satellites and lack of recent infall of new massive satellites over the last few Gyr. As a result, there should be an anti-correlation between the number of satellite galaxies and the mass of the central galaxy. Our estimate of correlation coefficient shows that there is a pretty significant anti-correlation, $r = -0.40 \pm 0.05$, between $N_{\rm gals}$ and $M_{\rm star, BCG}$. This anti-correlation implicitly supports the major merger scenario and can be used as another observable to test this hypothesis.

\subsection{Observational systematics and biases}

Mass of dark matter halos is not an observable; therefore, one has to rely on other observables to infer their mass. For example, to study the impact of the magnitude gap \citet{GoldenMarx:2018} relied on the caustic mass, \citet{Golden-Marx:2019} used optical-richness, and \citet{Hearin:2013} employed velocity dispersion as a proxy of the host halo mass. The correlation between these observables and other cluster properties can induce non-zero, change the value, or even flip the sign of the slope for the secondary explanatory variable. For example, the optical-richness is anti-correlation with M14 at fixed halo mass. As a result, we expect to see an anti-correlation between a property that is intrinsically uncorrelated with M14 but correlated with the optical-richness if the optical-richness is employed as a halo mass proxy.

Triaxiality and dynamical state of clusters are other sources of systematics that need special care. Triaxiality and dynamical state can induce bias in a sample selection process, and systematically affect the integrated measured quantities \citep{Dietrich:2014}. In signal-limited cluster samples, systems whose major axis points toward the observer are more likely to be included in the sample. Triaxiality can add bias in measured quantities, and under the standard assumption of spherical symmetry, the integrated quantities are over-estimated. This adds bias to the measured quantities.

Background, foreground, and interloper galaxies contaminate the optical observables by adding bais and, potentially correlated, noise to the measured quantities \citep{Costanzi:2018projection}. Some of the optical observables and explanatory variables are more susceptible to the projection effect. The X-ray and millimeter data, while cleaner than optical data, also suffer from the projection effect, which needs to be modeled and calibrated \citep{Noh:2012}. Therefore, extra caution should be exercised in interpreting the observational data and comparing them with simulations outputs. Before drawing a definite conclusion from observational data, first, the impact of the projection on these observables needs to be modeled and incorporated into a statistical analysis.


\section{Summary and Conclusion} \label{sec:conclusion}

Developing an accurate and precise halo mass estimator is an essential step in constraining the cosmological parameters with galaxy clusters. While cluster observable variance is studied extensively in the literature, it is one of the least constrained quantities and a source of systematic in a cluster cosmology analysis \citep{Farahi:2019, SDSS:2018cosmology}. Therefore, identifying and incorporating observables which can reduce the scatter in mass--observable relation is of particular interest.

The assembly history and dynamical interaction play out within the evolving cosmic web network of large-scale structure to determine the overall gas and stellar content of dark matter halos. We therefore speculate that observables which contain information about the assembly history of dark matter halos and their dynamical state can explain part of the scatter in HPHM relation. Thus, including these information in the scaling analysis enable us to predict the mass of these systems with higher precision. It has been suggested that the magnitude gap -- the magnitude difference between the brightest and the second or fourth brightest member galaxies -- probes the formation time of the cluster and can be employed to reduce the BCG stellar mass--halo mass relation \citep{GoldenMarx:2018}. Our goal is to study the impact of this observable in predicting the mass of the host halo.

In this work, we study the impact of incorporating the magnitude gap on the conditional statistics of halo properties, i.e., halo property--halo mass relation, by employing massive halos realized by TNG300 hydrodynamics simulations of the IllustrisTNG project. We summarize our key findings in the following.

\begin{itemize}
\item  Conditioning on the magnitude gap can efficiently reduce the scatter about the mean HPHM relation. The improvement in the scatter can be as large as $\sim 30\%$. While conditioning $\ln M_{\rm star}$ on magnitude gap does not reduce the scatter, conditioning $\ln M_{\rm star, BCG}$ or $\ln M_{\rm star, SAT}$ on magnitude gap can reduce scatter from $33\%$ to $\approx 23\%$. 
\item At fixed halo mass, larger(smaller) magnitude gap correspond to more(less) gas mass, more(less) massive BCG, smaller(larger) number of satellite galaxies with respect to the mean.  
\item Conditioning on the magnitude gap changes the halo property covariance. A halo property pair correlation can become stronger by removing the impact of mixing of halos of different age, or can become weaker if the correlation is mainly driven by the correlation with the formation time.
\item The log-normal model is sufficient for describing HPHM relation with/without conditioning on the magnitude gap.
\item We find that one can construct a better mass proxy, by splitting the integrated quantities into its components. For example, a linear relation of log of $M_{\rm star}$ and $M_{\rm gas}$ is a better estimator of halo mass than a linear relation of log of $M_{\rm b} = M_{\rm star} + M_{\rm gas}$. 
\end{itemize}

We provide qualitative and quantitative predictions on how the mean scaling relations and a property covariance response to conditioning on the magnitude gap. Current and future optical and near-infrared photometeric and spectroscopic surveys, such as DES, LSST, DESI, and WFIRST, can study these predictions and enhance our understanding of cluster formation and evolution. Preparing ourselves to interpret observational results and compare them with simulated halo quantities is a necessary step forward. For this next step, we need to develop an accurate and precise understanding of observational systematics and biases, such as the projection and selection effects. Therefore, a natural next step is to take simulated data into a light-cone, generate realistic mocked observations, and analyze the mocked observations.

\section*{Acknowledgments}

A. Farahi is supported by a McWilliams Postdoctoral Fellowship. H. Trac is pleased to acknowledge funding support from National Science Foundation grant IIS-1563887. The authors would like to thank August Evrard, Jesse Golden-Marx, Christopher Miller, Christopher Bradshaw and Michelle Ntampaka for useful comments and feedback. We thank the referee for their constructive feedback and criticism.  This work is performed in part at Aspen Center for Physics, which is supported by National Science Foundation grant PHY-1607611. We thank the IllustrisTNG team for making their data and catalogs publicly available.

\bibliographystyle{mn2e_adsurl}
\bibliography{apj-jour,astroref}

\begin{thebibliography}{}
  \providecommand{\doi}[1]{\href{http://dx.doi.org/#1}{doi:#1}}
  \providecommand{\eprint}[1]{\href{http://arxiv.org/abs/#1}{arXiv:#1}}

\bibitem[\protect\citeauthoryear{Bartlett}{Bartlett}{1951}]{bartlett1951}
\href{https://doi.org/10.1214/aoms/1177729698}{Bartlett M.~S.,  1951, Ann.
  Math. Statist., 22, 107}

\bibitem[\protect\citeauthoryear{Patil, Huard \& Fonnesbeck}{Patil
  et~al.}{2010}]{PyMC:2016}
Patil A.,  Huard D.,    Fonnesbeck C.~J.,  2010, Journal of statistical
  software, 35, 1

\bibitem[\protect\citeauthoryear{Sherman \& Morrison}{Sherman \&
  Morrison}{1950}]{sherman1950}
\href{https://doi.org/10.1214/aoms/1177729893}{Sherman J.,  Morrison W.~J.,
  1950, Ann. Math. Statist., 21, 124}

\bibitem[\protect\citeauthoryear{{Afshordi} \& {Cen}}{{Afshordi} \&
  {Cen}}{2002}]{Afshordi:2002}
\href{https://ui.adsabs.harvard.edu/abs/2002ApJ...564..669A}{{Afshordi} N.,
  {Cen} R.,  2002, \apj, 564, 669}

\bibitem[\protect\citeauthoryear{{Allen}, {Evrard} \& {Mantz}}{{Allen}
  et~al.}{2011}]{Allen:2011review}
\href{http://adsabs.harvard.edu/abs/2011ARA%26A..49..409A}{{Allen} S.~W.,
  {Evrard} A.~E.,    {Mantz} A.~B.,  2011, \araa, 49, 409}

\bibitem[\protect\citeauthoryear{{Anbajagane} et~al.,}{{Anbajagane}
  et~al.}{2020}]{Anbajagane:2020}
\href{https://ui.adsabs.harvard.edu/abs/2020arXiv200102283A}{{Anbajagane} D.
  et~al., 2020, arXiv e-prints, p. arXiv:2001.02283}

\bibitem[\protect\citeauthoryear{{Armitage}, {Kay} \& {Barnes}}{{Armitage}
  et~al.}{2019}]{Armitage:2019}
\href{https://ui.adsabs.harvard.edu/abs/2019MNRAS.484.1526A}{{Armitage} T.~J.,
  {Kay} S.~T.,    {Barnes} D.~J.,  2019, \mnras, 484, 1526}

\bibitem[\protect\citeauthoryear{{Behroozi} et~al.,}{{Behroozi}
  et~al.}{2019}]{Behroozi:2019}
\href{https://ui.adsabs.harvard.edu/abs/2019MNRAS.tmp.1134B}{{Behroozi} P.
  et~al., 2019, \mnras, p.~1134}

\bibitem[\protect\citeauthoryear{{Bertschinger}}{{Bertschinger}}{1985}]{Bertschinger:1985}
\href{http://adsabs.harvard.edu/abs/1985ApJS...58...39B}{{Bertschinger} E.,
  1985, Astrophys. J. Supp., 58, 39}

\bibitem[\protect\citeauthoryear{{Bradshaw} et~al.,}{{Bradshaw}
  et~al.}{2019}]{Bradshaw:2019}
\href{https://ui.adsabs.harvard.edu/abs/2019arXiv190509353B}{{Bradshaw} C.
  et~al., 2019, arXiv e-prints, p. arXiv:1905.09353}

\bibitem[\protect\citeauthoryear{{Cohn} \& {Battaglia}}{{Cohn} \&
  {Battaglia}}{2019}]{Cohn:2019}
\href{https://ui.adsabs.harvard.edu/abs/2019MNRAS.tmp.2686C}{{Cohn} J.~D.,
  {Battaglia} N.,  2019, \mnras, p.~2686}

\bibitem[\protect\citeauthoryear{{Cohn} et~al.,}{{Cohn}
  et~al.}{2007}]{Cohn:2007}
\href{https://ui.adsabs.harvard.edu/abs/2007MNRAS.382.1738C}{{Cohn} J.~D.
  et~al., 2007, \mnras, 382, 1738}

\bibitem[\protect\citeauthoryear{{Corsini} et~al.,}{{Corsini}
  et~al.}{2018}]{Corsini:2018}
\href{https://ui.adsabs.harvard.edu/abs/2018A%26A...618A.172C}{{Corsini} E.~M.
  et~al., 2018, \aap, 618, A172}

\bibitem[\protect\citeauthoryear{{Costanzi} et~al.,}{{Costanzi}
  et~al.}{2019a}]{Costanzi:2018projection}
\href{https://ui.adsabs.harvard.edu/\#abs/2019MNRAS.482..490C}{{Costanzi} M.
  et~al., 2019a, \mnras, 482, 490}

\bibitem[\protect\citeauthoryear{{Costanzi} et~al.,}{{Costanzi}
  et~al.}{2019b}]{SDSS:2018cosmology}
\href{https://ui.adsabs.harvard.edu/abs/2019MNRAS.488.4779C}{{Costanzi} M.
  et~al., 2019b, \mnras, 488, 4779}

\bibitem[\protect\citeauthoryear{{D'Onghia} et~al.,}{{D'Onghia}
  et~al.}{2005}]{DOnghia:2005}
\href{https://ui.adsabs.harvard.edu/abs/2005ApJ...630L.109D}{{D'Onghia} E.
  et~al., 2005, \apjl, 630, L109}

\bibitem[\protect\citeauthoryear{{Dariush} et~al.,}{{Dariush}
  et~al.}{2010}]{Dariush:2010}
\href{https://ui.adsabs.harvard.edu/abs/2010MNRAS.405.1873D}{{Dariush} A.~A.
  et~al., 2010, \mnras, 405, 1873}

\bibitem[\protect\citeauthoryear{{Deason} et~al.,}{{Deason}
  et~al.}{2013}]{Deason:2013}
\href{https://ui.adsabs.harvard.edu/abs/2013ApJ...777..154D}{{Deason} A.~J.
  et~al., 2013, \apj, 777, 154}

\bibitem[\protect\citeauthoryear{{Dietrich} et~al.,}{{Dietrich}
  et~al.}{2014}]{Dietrich:2014}
\href{https://ui.adsabs.harvard.edu/abs/2014MNRAS.443.1713D}{{Dietrich} J.~P.
  et~al., 2014, \mnras, 443, 1713}

\bibitem[\protect\citeauthoryear{{Erickson}, {Cunha} \& {Evrard}}{{Erickson}
  et~al.}{2011}]{Erickson:2011}
\href{http://adsabs.harvard.edu/abs/2011PhRvD..84j3506E}{{Erickson} B.~M.~S.,
  {Cunha} C.~E.,    {Evrard} A.~E.,  2011, \prd, 84, 103506}

\bibitem[\protect\citeauthoryear{{Evrard} et~al.,}{{Evrard}
  et~al.}{2002}]{Evrard:2002}
\href{http://adsabs.harvard.edu/abs/2002ApJ...573....7E}{{Evrard} A.~E.
  et~al., 2002, \apj, 573, 7}

\bibitem[\protect\citeauthoryear{{Evrard} et~al.,}{{Evrard}
  et~al.}{2008}]{Evrard:2008}
\href{http://adsabs.harvard.edu/abs/2008ApJ...672..122E}{{Evrard} A.~E.
  et~al., 2008, \apj, 672, 122}

\bibitem[\protect\citeauthoryear{{Evrard} et~al.,}{{Evrard}
  et~al.}{2014}]{Evrard:2014}
\href{http://adsabs.harvard.edu/abs/2014MNRAS.441.3562E}{{Evrard} A.~E.
  et~al., 2014, \mnras, 441, 3562}

\bibitem[\protect\citeauthoryear{{Farahi} et~al.,}{{Farahi}
  et~al.}{2018}]{Farahi:2018bahamas}
\href{http://adsabs.harvard.edu/abs/2017arXiv171104922F}{{Farahi} A.  et~al.,
  2018, \mnras, 478, 2618}

\bibitem[\protect\citeauthoryear{{Farahi} et~al.,}{{Farahi}
  et~al.}{2019a}]{Farahi:2019anti}
\href{https://ui.adsabs.harvard.edu/abs/2019NatCo..10.2504F}{{Farahi} A.
  et~al., 2019a, Nature Communications, 10, 2504}

\bibitem[\protect\citeauthoryear{{Farahi} et~al.,}{{Farahi}
  et~al.}{2019b}]{Farahi:2019}
\href{https://ui.adsabs.harvard.edu/abs/2019MNRAS.490.3341F}{{Farahi} A.
  et~al., 2019b, \mnras, 490, 3341}

\bibitem[\protect\citeauthoryear{{Fujita} et~al.,}{{Fujita}
  et~al.}{2018}]{Fujita:2018}
\href{https://ui.adsabs.harvard.edu/abs/2018ApJ...857..118F}{{Fujita} Y.
  et~al., 2018, \apj, 857, 118}

\bibitem[\protect\citeauthoryear{{Golden-Marx} \& {Miller}}{{Golden-Marx} \&
  {Miller}}{2018}]{GoldenMarx:2018}
\href{https://ui.adsabs.harvard.edu/abs/2018ApJ...860....2G}{{Golden-Marx}
  J.~B.,  {Miller} C.~J.,  2018, \apj, 860, 2}

\bibitem[\protect\citeauthoryear{{Golden-Marx} \& {Miller}}{{Golden-Marx} \&
  {Miller}}{2019}]{Golden-Marx:2019}
\href{https://ui.adsabs.harvard.edu/abs/2019ApJ...878...14G}{{Golden-Marx}
  J.~B.,  {Miller} C.~J.,  2019, \apj, 878, 14}

\bibitem[\protect\citeauthoryear{{Green} et~al.,}{{Green}
  et~al.}{2019}]{Green:2019}
\href{https://ui.adsabs.harvard.edu/abs/2019ApJ...884...33G}{{Green} S.~B.
  et~al., 2019, \apj, 884, 33}

\bibitem[\protect\citeauthoryear{{Hartley} et~al.,}{{Hartley}
  et~al.}{2008}]{Hartley:2008}
\href{https://ui.adsabs.harvard.edu/abs/2008MNRAS.386.2015H}{{Hartley} W.~G.
  et~al., 2008, \mnras, 386, 2015}

\bibitem[\protect\citeauthoryear{{Hearin} et~al.,}{{Hearin}
  et~al.}{2013}]{Hearin:2013}
\href{https://ui.adsabs.harvard.edu/abs/2013MNRAS.430.1238H}{{Hearin} A.~P.
  et~al., 2013, \mnras, 430, 1238}

\bibitem[\protect\citeauthoryear{{Ho} et~al.,}{{Ho} et~al.}{2019}]{Ho:2019}
\href{https://ui.adsabs.harvard.edu/abs/2019ApJ...887...25H}{{Ho} M.  et~al.,
  2019, \apj, 887, 25}

\bibitem[\protect\citeauthoryear{{Huterer} \& {Shafer}}{{Huterer} \&
  {Shafer}}{2018}]{Huterer:2018}
\href{http://adsabs.harvard.edu/abs/2018RPPh...81a6901H}{{Huterer} D.,
  {Shafer} D.~L.,  2018, Reports on Progress in Physics, 81, 016901}

\bibitem[\protect\citeauthoryear{{Jenkins} et~al.,}{{Jenkins}
  et~al.}{2001}]{Jenkins:2001}
\href{http://adsabs.harvard.edu/abs/2001MNRAS.321..372J}{{Jenkins} A.  et~al.,
  2001, \mnras, 321, 372}

\bibitem[\protect\citeauthoryear{{Jiang} \& {van den Bosch}}{{Jiang} \& {van
  den Bosch}}{2017}]{Jiang:2017}
\href{https://ui.adsabs.harvard.edu/abs/2017MNRAS.472..657J}{{Jiang} F.,  {van
  den Bosch} F.~C.,  2017, \mnras, 472, 657}

\bibitem[\protect\citeauthoryear{{Jones} et~al.,}{{Jones}
  et~al.}{2003}]{Jones:2003}
\href{https://ui.adsabs.harvard.edu/abs/2003MNRAS.343..627J}{{Jones} L.~R.
  et~al., 2003, \mnras, 343, 627}

\bibitem[\protect\citeauthoryear{{Kaiser}}{{Kaiser}}{1991}]{Kaiser:1991}
\href{http://adsabs.harvard.edu/abs/1991ApJ...383..104K}{{Kaiser} N.,  1991,
  \apj, 383, 104}

\bibitem[\protect\citeauthoryear{{Kundert}, {D'Onghia} \& {Aguerri}}{{Kundert}
  et~al.}{2017}]{Kundert:2017}
\href{https://ui.adsabs.harvard.edu/abs/2017ApJ...845...45K}{{Kundert} A.,
  {D'Onghia} E.,    {Aguerri} J.~A.~L.,  2017, \apj, 845, 45}

\bibitem[\protect\citeauthoryear{{Larson}}{{Larson}}{1973}]{Larson:1973}
\href{https://ui.adsabs.harvard.edu/abs/1973MNRAS.161..133L}{{Larson} R.~B.,
  1973, \mnras, 161, 133}

\bibitem[\protect\citeauthoryear{{Le Brun} et~al.,}{{Le Brun}
  et~al.}{2017}]{LeBrun:2017}
\href{http://adsabs.harvard.edu/abs/2017MNRAS.466.4442L}{{Le Brun} A.~M.~C.
  et~al., 2017, \mnras, 466, 4442}

\bibitem[\protect\citeauthoryear{{Lieu} et~al.,}{{Lieu}
  et~al.}{2016}]{Lieu:2016xxl}
\href{https://ui.adsabs.harvard.edu/\#abs/2016A&A...592A...4L}{{Lieu} M.
  et~al., 2016, \aap, 592, A4}

\bibitem[\protect\citeauthoryear{{Lynden-Bell}}{{Lynden-Bell}}{1967}]{LyndenBell:1967}
\href{https://ui.adsabs.harvard.edu/abs/1967MNRAS.136..101L}{{Lynden-Bell} D.,
  1967, \mnras, 136, 101}

\bibitem[\protect\citeauthoryear{{Mahdavi} et~al.,}{{Mahdavi}
  et~al.}{2013}]{Mahdavi:2013}
\href{http://adsabs.harvard.edu/abs/2013ApJ...767..116M}{{Mahdavi} A.  et~al.,
  2013, \apj, 767, 116}

\bibitem[\protect\citeauthoryear{{Mantz} et~al.,}{{Mantz}
  et~al.}{2010}]{Mantz:2010}
\href{http://adsabs.harvard.edu/abs/2010MNRAS.406.1773M}{{Mantz} A.  et~al.,
  2010, \mnras, 406, 1773}

\bibitem[\protect\citeauthoryear{{Mantz} et~al.,}{{Mantz}
  et~al.}{2014}]{Mantz:2014fb}
\href{https://ui.adsabs.harvard.edu/abs/2014MNRAS.440.2077M}{{Mantz} A.~B.
  et~al., 2014, \mnras, 440, 2077}

\bibitem[\protect\citeauthoryear{{Mantz} et~al.,}{{Mantz}
  et~al.}{2016a}]{Mantz:2016-relaxedIII}
\href{http://adsabs.harvard.edu/abs/2016MNRAS.456.4020M}{{Mantz} A.~B.  et~al.,
  2016a, \mnras, 456, 4020}

\bibitem[\protect\citeauthoryear{{Mantz} et~al.,}{{Mantz}
  et~al.}{2016b}]{Mantz:2016WtG-V}
\href{http://adsabs.harvard.edu/abs/2016MNRAS.463.3582M}{{Mantz} A.~B.  et~al.,
  2016b, \mnras, 463, 3582}

\bibitem[\protect\citeauthoryear{{Marinacci} et~al.,}{{Marinacci}
  et~al.}{2018}]{Marinacci:2018}
\href{https://ui.adsabs.harvard.edu/abs/2018MNRAS.480.5113M}{{Marinacci} F.
  et~al., 2018, \mnras, 480, 5113}

\bibitem[\protect\citeauthoryear{{Matthee} et~al.,}{{Matthee}
  et~al.}{2017}]{Matthee:2017}
\href{https://ui.adsabs.harvard.edu/abs/2017MNRAS.465.2381M}{{Matthee} J.
  et~al., 2017, \mnras, 465, 2381}

\bibitem[\protect\citeauthoryear{{McCarthy} et~al.,}{{McCarthy}
  et~al.}{2017}]{McCarthy:2017}
\href{https://ui.adsabs.harvard.edu/abs/2017MNRAS.465.2936M}{{McCarthy} I.~G.
  et~al., 2017, \mnras, 465, 2936}

\bibitem[\protect\citeauthoryear{{McClintock} et~al.,}{{McClintock}
  et~al.}{2019}]{McClintock:2019}
\href{https://ui.adsabs.harvard.edu/abs/2019MNRAS.482.1352M}{{McClintock} T.
  et~al., 2019, \mnras, 482, 1352}

\bibitem[\protect\citeauthoryear{{Mulroy} et~al.,}{{Mulroy}
  et~al.}{2019}]{Mulroy:2019}
\href{https://ui.adsabs.harvard.edu/abs/2019MNRAS.484...60M}{{Mulroy} S.~L.
  et~al., 2019, \mnras, 484, 60}

\bibitem[\protect\citeauthoryear{{Murray}, {Power} \& {Robotham}}{{Murray}
  et~al.}{2013}]{Murray:2013}
\href{http://adsabs.harvard.edu/abs/2013A%26C.....3...23M}{{Murray} S.~G.,
  {Power} C.,    {Robotham} A.~S.~G.,  2013, Astronomy and Computing, 3, 23}

\bibitem[\protect\citeauthoryear{{Naiman} et~al.,}{{Naiman}
  et~al.}{2018}]{Naiman:2018}
\href{https://ui.adsabs.harvard.edu/abs/2018MNRAS.477.1206N}{{Naiman} J.~P.
  et~al., 2018, \mnras, 477, 1206}

\bibitem[\protect\citeauthoryear{{Nelson} et~al.,}{{Nelson}
  et~al.}{2018}]{Nelson:2018}
\href{https://ui.adsabs.harvard.edu/abs/2018MNRAS.475..624N}{{Nelson} D.
  et~al., 2018, \mnras, 475, 624}

\bibitem[\protect\citeauthoryear{{Noh} \& {Cohn}}{{Noh} \&
  {Cohn}}{2012}]{Noh:2012}
\href{https://ui.adsabs.harvard.edu/abs/2012MNRAS.426.1829N}{{Noh} Y.,  {Cohn}
  J.~D.,  2012, \mnras, 426, 1829}

\bibitem[\protect\citeauthoryear{{Ntampaka} et~al.,}{{Ntampaka}
  et~al.}{2015}]{Ntampaka:2015}
\href{https://ui.adsabs.harvard.edu/abs/2015ApJ...803...50N}{{Ntampaka} M.
  et~al., 2015, \apj, 803, 50}

\bibitem[\protect\citeauthoryear{{Ntampaka} et~al.,}{{Ntampaka}
  et~al.}{2016}]{Ntampaka:2016}
\href{https://ui.adsabs.harvard.edu/abs/2016ApJ...831..135N}{{Ntampaka} M.
  et~al., 2016, \apj, 831, 135}

\bibitem[\protect\citeauthoryear{{Ntampaka} et~al.,}{{Ntampaka}
  et~al.}{2019a}]{Ntampaka:2019MLinAstro}
\href{https://ui.adsabs.harvard.edu/abs/2019BAAS...51c..14N}{{Ntampaka} M.
  et~al., 2019a, BAAS, 51, 14}

\bibitem[\protect\citeauthoryear{{Ntampaka} et~al.,}{{Ntampaka}
  et~al.}{2019b}]{Ntampaka:2019}
\href{https://ui.adsabs.harvard.edu/abs/2019ApJ...876...82N}{{Ntampaka} M.
  et~al., 2019b, \apj, 876, 82}

\bibitem[\protect\citeauthoryear{{Okoli} \& {Afshordi}}{{Okoli} \&
  {Afshordi}}{2016}]{Okoli:2016}
\href{https://ui.adsabs.harvard.edu/abs/2016MNRAS.456.3068O}{{Okoli} C.,
  {Afshordi} N.,  2016, \mnras, 456, 3068}

\bibitem[\protect\citeauthoryear{{Palmese} et~al.,}{{Palmese}
  et~al.}{2019}]{Palmese:2019}
\href{https://ui.adsabs.harvard.edu/abs/2019arXiv190308813P}{{Palmese} A.
  et~al., 2019, arXiv e-prints, p. arXiv:1903.08813}

\bibitem[\protect\citeauthoryear{{Parekh} et~al.,}{{Parekh}
  et~al.}{2015}]{Parekh:2015}
\href{https://ui.adsabs.harvard.edu/abs/2015A&A...575A.127P}{{Parekh} V.
  et~al., 2015, \aap, 575, A127}

\bibitem[\protect\citeauthoryear{{Pillepich} et~al.,}{{Pillepich}
  et~al.}{2018a}]{Pillepich:2018}
\href{https://ui.adsabs.harvard.edu/abs/2018MNRAS.473.4077P}{{Pillepich} A.
  et~al., 2018a, \mnras, 473, 4077}

\bibitem[\protect\citeauthoryear{{Pillepich} et~al.,}{{Pillepich}
  et~al.}{2018b}]{Pillepich:2018scaling}
\href{https://ui.adsabs.harvard.edu/abs/2018MNRAS.475..648P}{{Pillepich} A.
  et~al., 2018b, \mnras, 475, 648}

\bibitem[\protect\citeauthoryear{{Ponman} et~al.,}{{Ponman}
  et~al.}{1994}]{Ponman:1994}
\href{https://ui.adsabs.harvard.edu/abs/1994Natur.369..462P}{{Ponman} T.~J.
  et~al., 1994, Nature, 369, 462}

\bibitem[\protect\citeauthoryear{{Pratt} et~al.,}{{Pratt}
  et~al.}{2009}]{Pratt:2009}
\href{http://adsabs.harvard.edu/abs/2009A%26A...498..361P}{{Pratt} G.~W.
  et~al., 2009, \aap, 498, 361}

\bibitem[\protect\citeauthoryear{{Reichert} et~al.,}{{Reichert}
  et~al.}{2011}]{Reichert:2011}
\href{http://adsabs.harvard.edu/abs/2011A%26A...535A...4R}{{Reichert} A.
  et~al., 2011, \aap, 535, A4}

\bibitem[\protect\citeauthoryear{{Rykoff} et~al.,}{{Rykoff}
  et~al.}{2016}]{Rykoff:2016}
\href{http://adsabs.harvard.edu/abs/2016ApJS..224....1R}{{Rykoff} E.~S.
  et~al., 2016, \apjs, 224, 1}

\bibitem[\protect\citeauthoryear{{Shaw}, {Holder} \& {Dudley}}{{Shaw}
  et~al.}{2010}]{Shaw:2010}
\href{http://adsabs.harvard.edu/abs/2010ApJ...716..281S}{{Shaw} L.~D.,
  {Holder} G.~P.,    {Dudley} J.,  2010, \apj, 716, 281}

\bibitem[\protect\citeauthoryear{{Springel}}{{Springel}}{2010}]{Springel:2010EMesh}
\href{https://ui.adsabs.harvard.edu/abs/2010MNRAS.401..791S}{{Springel} V.,
  2010, \mnras, 401, 791}

\bibitem[\protect\citeauthoryear{{Springel} et~al.,}{{Springel}
  et~al.}{2018}]{Springel:2018}
\href{https://ui.adsabs.harvard.edu/abs/2018MNRAS.475..676S}{{Springel} V.
  et~al., 2018, \mnras, 475, 676}

\bibitem[\protect\citeauthoryear{{Stanek} et~al.,}{{Stanek}
  et~al.}{2010}]{Stanek:2010}
\href{http://adsabs.harvard.edu/abs/2010ApJ...715.1508S}{{Stanek} R.  et~al.,
  2010, \apj, 715, 1508}

\bibitem[\protect\citeauthoryear{{Tinker} et~al.,}{{Tinker}
  et~al.}{2008}]{Tinker:2008}
\href{http://adsabs.harvard.edu/abs/2008ApJ...688..709T}{{Tinker} J.  et~al.,
  2008, \apj, 688, 709}

\bibitem[\protect\citeauthoryear{{Truong} et~al.,}{{Truong}
  et~al.}{2018}]{Truong:2018}
\href{http://adsabs.harvard.edu/abs/2018MNRAS.474.4089T}{{Truong} N.  et~al.,
  2018, \mnras, 474, 4089}

\bibitem[\protect\citeauthoryear{{Vikhlinin} et~al.,}{{Vikhlinin}
  et~al.}{2009}]{Vikhlinin:2009scaling}
\href{https://ui.adsabs.harvard.edu/abs/2009ApJ...692.1033V}{{Vikhlinin} A.
  et~al., 2009, \apj, 692, 1033}

\bibitem[\protect\citeauthoryear{{Vitorelli} et~al.,}{{Vitorelli}
  et~al.}{2018}]{Vitorelli:2018}
\href{https://ui.adsabs.harvard.edu/abs/2018MNRAS.474..866V}{{Vitorelli} A.~Z.
  et~al., 2018, \mnras, 474, 866}

\bibitem[\protect\citeauthoryear{{Vogelsberger} et~al.,}{{Vogelsberger}
  et~al.}{2014}]{Vogelsberger:2014}
\href{https://ui.adsabs.harvard.edu/abs/2014MNRAS.444.1518V}{{Vogelsberger} M.
  et~al., 2014, \mnras, 444, 1518}

\bibitem[\protect\citeauthoryear{{Wechsler} \& {Tinker}}{{Wechsler} \&
  {Tinker}}{2018}]{Wechsler:2018}
\href{https://ui.adsabs.harvard.edu/abs/2018ARA&A..56..435W}{{Wechsler} R.~H.,
  {Tinker} J.~L.,  2018, \araa, 56, 435}

\bibitem[\protect\citeauthoryear{{Wechsler} et~al.,}{{Wechsler}
  et~al.}{2002}]{Wechsler:2002}
\href{https://ui.adsabs.harvard.edu/abs/2002ApJ...568...52W}{{Wechsler} R.~H.
  et~al., 2002, \apj, 568, 52}

\bibitem[\protect\citeauthoryear{{Weinberger}, {Springel} \&
  {Pakmor}}{{Weinberger} et~al.}{2019}]{Weinberger:2019}
\href{https://ui.adsabs.harvard.edu/abs/2019arXiv190904667W}{{Weinberger} R.,
  {Springel} V.,    {Pakmor} R.,  2019, arXiv e-prints, p. arXiv:1909.04667}

\bibitem[\protect\citeauthoryear{{Weinberg} et~al.,}{{Weinberg}
  et~al.}{2013}]{Weinberg:2013}
\href{http://adsabs.harvard.edu/abs/2013PhR...530...87W}{{Weinberg} D.~H.
  et~al., 2013, \physrep, 530, 87}

\bibitem[\protect\citeauthoryear{{Wu} et~al.,}{{Wu} et~al.}{2013}]{Wu:2013}
\href{https://ui.adsabs.harvard.edu/abs/2013ApJ...767...23W}{{Wu} H.-Y.
  et~al., 2013, \apj, 767, 23}

\bibitem[\protect\citeauthoryear{{Wu} et~al.,}{{Wu} et~al.}{2015}]{Wu:2015}
\href{https://ui.adsabs.harvard.edu/abs/2015MNRAS.452.1982W}{{Wu} H.-Y.
  et~al., 2015, \mnras, 452, 1982}

\bibitem[\protect\citeauthoryear{{Zinnecker}}{{Zinnecker}}{1984}]{Zinnecker:1984}
\href{https://ui.adsabs.harvard.edu/abs/1984MNRAS.210...43Z}{{Zinnecker} H.,
  1984, \mnras, 210, 43}

\bibitem[\protect\citeauthoryear{{van den Bosch} \& {Ogiya}}{{van den Bosch} \&
  {Ogiya}}{2018}]{vandenBosch:2018}
\href{https://ui.adsabs.harvard.edu/abs/2018MNRAS.475.4066V}{{van den Bosch}
  F.~C.,  {Ogiya} G.,  2018, \mnras, 475, 4066}

\end{thebibliography}

\appendix

\section{Splitting the intrinsic scatter covariance matrix} \label{app:covariance-splitting}

 In the main text, we show the impact of conditioning on a secondary explanatory variable on the covariance matrix and reducing the scatter. Here, we provide a simple mathematical model which allows us to capture and quantify this effect. The likelihood of a set of halo properties at fixed halo mass can be computed by marginalizing over the secondary parameter,
\begin{equation} \label{eq:app-model}
    P(\textbf{s}_p\, |\, M) = \int\, {\rm d}X_2 \, P(\textbf{s}_p\, |\, X_2, M) \, P(X_2\, |\, M).
\end{equation}
Not conditioning on the secondary variable is mathematically equivalent to marginalizing over it. We assume log halo properties conditioned on halo mass and $X_2$ can be modeled with a multivariate normal distribution. We further assume $P(X_2\, |\, M)$ follows a normal distribution with mean zero and variance $\sigma_X^2$. $P(\textbf{s}_p\, |\, M)$ also becomes a multi-variate normal distribution under the current assumptions. The log-mean relation linearly runs with log mass and $X_2$
\begin{equation} 
 \langle s_{a} \, | \, M, X_2 \rangle \ = \ 
  \pi_a  + \alpha_a \, \ln M_{\rm halo} + \beta_a \, X_2\,, 
\end{equation}
where $\pi_a$ is the normalization and $\alpha_a$ and $\beta_a$ are the slopes of log mass and the secondary variable, respectively. For brevity, we denote ${\rm COV}(\textbf{s}_p  \,|\, M, X_2)$ with $C$ and ${\rm COV}(\textbf{s}_p  \,|\, M)$ with $C_m$. We refer to $C$ and $C_m$ as intrinsic and marginalized covarinaces. The Gaussian assumptions allow us to analytically derive the marginalized halo property covariance by performing Equation (\ref{eq:app-model}):
\begin{equation}
    C^{-1}_m = C^{-1} - \frac{C^{-1} \pmb{\beta} \pmb{\beta}^T  C^{-1} }{1 + \sigma_X^2 \, \pmb{\beta}^T C^{-1} \pmb{\beta} } \sigma_X^2 \,.
\end{equation}
We employ Sherman-Morrison identity \citep{sherman1950, bartlett1951} to compute $C_m$, 
\begin{equation} \label{eq:app-margin-covariance}
    C_m = C + \pmb{\beta} \pmb{\beta}^T \sigma_X^2 \,.
\end{equation}
The second term in Equation (\ref{eq:app-margin-covariance}) quantifies how much information is lost due to marginalizing over $X_2$. If halo properties are not correlated with the secondary parameters, i.e. $\pmb{\beta} = [0 \cdots 0]_{N \times 1}$, then the second term becomes zero which implies $C_m = C$, as expected. 

For simplicity, let us consider a vector of halo properties with two properties. The marginalized and intrinsic covariance matrices each have three free parameters, two variances and a correlation coefficient. We denote the intrinsic property pair correlation with $r_{\rm int}$ and the marginalized property pair correlation with $r$. The scenario in which property $a$ is correlated with the secondary property while property $b$ is uncorrelated can be modeled with $\pmb{\beta} = [\beta_a \,, 0]$ where $\beta_a \neq 0$. The marginalized covariance becomes 
\begin{equation*}
C_m  = C +
\begin{pmatrix}
   \sigma^2_{X} \, \beta_a^2 & 0  \\
    0 & 0  \\
\end{pmatrix} \,,
\end{equation*}
This leads to the suppression of the marginalized property pair correlation between property $a$ and $b$ at fixed halo mass. The relation between $r$ and $r_{\rm int}$ becomes
\begin{equation}
   r = r_{\rm int} \frac{\sigma^2_{a}}{\sigma^2_{a} + \sigma^2_{X} \beta_a^2}\,.
\end{equation}
In conclusion, when one quantity is uncorrelated with the secondary variable, marginalizing over the secondary variable suppresses the measured property pair correlation, i.e. $| r | < | r_{\rm int} |$. This model explains why the correlation coefficient between $M_{\rm star}$ and other halo properties when they are conditioned on $\ln M$ and $M14$ is always stronger than property pair correlation with they are only conditioned on $\ln M$.

\label{lastpage}

\end{document}